\documentstyle[eqsecnum,pra,aps]{revtex}

\begin{document}

\title{Quasi-particle excitations and dynamical structure
function of trapped Bose-condensates in
the WKB approximation}

\author{Andr\'as Csord\'as${}^{*}$, Robert Graham${}^{\dagger}$, and
P\'eter Sz\'epfalusy${}^{**\dagger\dagger}$} %
\maketitle
\centerline{${}^{*}$Research Group for Statistical Physics of the
Hungarian Academy of Sciences,}
\centerline{ M\'uzeum krt. 6--8, H-1088 Budapest,
Hungary}
\centerline{${}^{\dagger}$Fachbereich Physik,
Universit\"at-Gesamthochschule Essen,}
\centerline{45117 Essen,Germany}

\centerline{${}^{**}$Institute for Solid State
Physics, E\"otv\"os University,}
\centerline{ M\'uzeum krt. 6--8, H-1088 Budapest,
Hungary}

\centerline{${}^{\dagger\dagger}$Research Institute
for Solid State Physics, P.O. Box 49, H--1525 Budapest, Hungary,}

\begin{abstract}
The Bogoliubov equations of the quasi-particle excitations in
a weakly interacting trapped Bose-condensate are solved in the WKB
approximation in an isotropic harmonic trap, determining the discrete  
quasi-particle energies and wave functions by torus (Bohr-Sommerfeld) quantization
of the integrable classical quasi-particle dynamics.
The results are used to calculate the position and strengths of the
peaks in the dynamic structure function which can be observed by
off-resonance inelastic light-scattering.
\end{abstract}


\section{Introduction}
\label{sec:intro}
The experimental realization of Bose-Einstein condensates of weakly
interacting alkali atoms confined by magnetic traps \cite{1a,2a,3a,4a} has revived the
interest in the properties of Bose-condensed systems \cite {11a}, in particular in solutions of the Gross-Pitaevski \cite{5aa} equation in an external
potential, describing the condensate, and of the Bogoliubov
equations governing the quasi-particle excitations out of the spatially
inhomogeneous condensate \cite{6,6a,6b}. Approximate spatially inhomogeneous solutions
of the Gross-Pitaevski equation have been obtained by
a Gaussian variational ansatz, which is appropriate for moderate size
condensates, and  by the Thomas-Fermi 
approximation \cite{7a}, which is very good
for large condensates, except in a narrow region around the surface of
the condensate. Similarly, solutions of the Bogoliubov equations
for the low-lying quasi-particle excitations have been constructed
numerically \cite{6}, by a variational method \cite{6a}, and in the hydrodynamic limit \cite{6b}.
Such low-lying states have recently been excited by resonant periodic
variations in the trap parameters \cite{4,5,60}, and good agreement with the
theoretical predictions has been obtained. 

Much less work has been done
so far on the high lying excitations, where all the methods which have
so far been used for the low lying ones are either impracticable or
don't apply. Such high lying states are of experimental interest because
they are excited by off-resonance inelastic light
scattering. In
a previous paper \cite{Csordas1} we have computed the high-lying quasi-particle
energies and the amplitude of the inelastic peaks in the
scattering function using the Thomas-Fermi approximation and
solving the Hartree-Fock-Bogoliubov equations in perturbation theory of first
order for the energies (with an estimate of the second order ones) 
but only to zero order in the wave functions.
It is desirable to improve on this highly simplified approach by developing
a self-consistent non-perturbative approach and extend the calculation
to the whole energy region. 
It is our purpose, in the present paper, to present
a non-perturbative method --- the WKB approximation --- for the
solution of the Bogoliubov equations in a trap which allows us to interpolate  
between the hydrodynamical and the high-energy regime, and to use it to  
calculate the energies
and wave-functions of the quasi-particle states. In a previous brief report we have  already presented some results of
our calculation of the high lying levels of the quasi-particles \cite{Csordas2}. Here we shall apply our results to calculate the dynamical structure function, which is of
high interest, because it could be measured in off-resonance light scattering
\cite{Shl,Pol,Jav}. Calculations of the dynamical structure function have previously been presented for the homogeneous ideal Bose gas \cite{Jav,Ruo} and for the homogeneous interacting Bose gas \cite{Gr}, which are however not directly applicable to the trapped condensates. Results have also been presented for the hydrodynamic limit of the structure function \cite{Wu}.
Very recently results for the dynamical structure function in the local density approximation have been presented \cite{Tom}, However, due to the limitations of that approximation, the resulting structure function cannot show structure beyond the constraints imposed by the locally averaged energy and momentum conservation.

The results we shall present here are applicable in a large enery domain ranging from low lying to high lying levels of the trapped condensates and  retain the full information on the discrete nature of the quasi-particle levels. They are,however, restricted to the spatially isotropic case. In previous work \cite{Csordas1} we have calculated the dynamical structure function for the very high lying quasiparticle levels in isotropic trapped condensates in a simpler approximation, replacing the high lying states by free trap states. The more sophisticated non-perturbative calculation presented in the present paper shows, however, 
that the interaction of the high lying states with the condensate has a non-negligible influence on the selection of the angular momentum of the
states excited by light-scattering.

An essential and independent input for our WKB-calculations is the
condensate wave-function, which must be obtained by solving the
Gross-Pitaevski equation. As we intend to carry out the WKB calculations
analytically, as far as possible, we shall use here the main analytical
approximation for large Bose-condensates with repulsive interaction,
the Thomas-Fermi approximation. 
The necessary results are collected in section~\ref{sec:bas}.

In section~\ref{sec:WKB}.\ the WKB approximation for the 
Boguliubov equations in an
isotropic harmonic trap is
developed. It is implemented in section~\ref{sec:bohr}.\ by studying
the classical quasi-particle dynamics, which is integrable in isotropic traps.
For a discussion of the non-integrable quasi-particle dynamics in anisotropic traps see \cite{Fliesser2}.

The reduced action is calculated as a function of
energy and angular momentum. In section~\ref{sec:bohr}.\ the Bohr-Sommerfeld quantization
of the energy levels of the classical quasi-particle dynamics is performed
and the semiclassical wave functions, properly normalized, are obtained in
section~\ref{sec:waves}.
The necessary integrals, though tedious, can all be done analytically.

In section~\ref{sec:scattering}.\ we present
our results for the scattering function and in section~\ref{sec:numerics}
some numerical examples are given. 
Our conclusions are summarized in
the final section~\ref{sec:conclusion}.

\section{Basic equations for condensate and quasi-particle excitations}
\label{sec:bas}
A necessary prerequisite of any calculation of the quasi-particle
excitations is an appropriate description of the condensate. The relevant
results are summarized in the present section. For weakly interacting atoms
at zero temperature practically all particles are in the condensate and the  
energy $E$ is a functional of the macroscopic condensate wave function  
$\psi_0(\bbox{r})$
 \begin{equation}
  E(\{\psi_0\})=\int d^2r\left[\frac{\hbar^2}{2m}
   |\bbox{\nabla}\psi_0(\bbox{r})|^2+U(\bbox{r})|\psi_0(\bbox{r})|^2
   +\frac{2\pi\hbar^2a}{m}
   |\psi_0(\bbox{r})|^4\right]
  \label{eq:2.1}
 \end{equation}
where the constraint
 \begin{equation}
  \int d^3r|\psi_0(\bbox{r})|^2=N_0
  \label{eq:2.2}
 \end{equation}
must be imposed. $U(\bbox{r})$ is the trap potential. In our calculations we
shall assume an isotropic harmonic trap
 \begin{equation}
  U(r)=\frac{1}{2}m\omega^2_0r^2
  \label{eq:2.3}
 \end{equation}
for simplicity. The scattering length is assumed to be positive and is
denoted by $a$. The number of atoms in the condensate is denoted by $N_0$.
We assume that the interaction is very weak and the temperature is close
to zero, so that $N_0\approx N$. The minimization condition for $E$,  
introducing the chemical potential $\mu$ as a Lagrange
multiplier to enforce the constraint (\ref{eq:2.2}), yields the
Gross-Pitaevski equation \cite{5aa}
\begin{equation}
 -\frac{\hbar^2}{2m}\nabla^2\psi_0+U(\bbox{r})\psi_0+
  \frac{4\pi\hbar^2 a}{m}|\psi_0|^2\psi_0=\mu\psi_0\,.
\label{eq:2.3a}
\end{equation}

\subsection{Thomas-Fermi approximation}
In the present paper we shall use as an analytical approximation for
the minimum of (\ref{eq:2.1}) the Thomas Fermi approximation, which holds in  
the limit of large condensates
$N_0a>>d_0\equiv\sqrt{\frac{\hbar}{m\omega_0}}$. In that limit it becomes a  
very good approximation to solve (\ref{eq:2.3a}) by neglecting the kinetic  
energy term \cite{7a}
\begin{equation}
 \psi_{0}(r)=\left(\frac{m}{4\pi\hbar^2a}
  (\mu-U(r))\right)^{1/2}\Theta(\mu-U(r))
\label{eq:2.4}
\end{equation}
where $\mu$ must be determined as a function of $N_0$ by imposing the
normalization condition (\ref{eq:2.2}), which yields
\begin{equation}
 \mu=\frac{\hbar\omega_0}{2}(15N_0a/d_0)^{2/5}
\label{eq:2.4d}
\end{equation}
The conditions $\mu>>\hbar\omega_0$ and $N_0a>>d_0$ are therefore equivalent.
In eq.~(\ref{eq:2.4}) $\Theta(x)$
is the step function. The approximation eq.~(\ref{eq:2.4}) with discontinuous  
derivative at
$r=r_0=\sqrt{2\mu/m\omega_0^2}$ is sufficient, except in those cases 
where derivatives of the
wave-function $\psi_0(r)$ for $r$ close to $r_0$ are important. In the latter  
case a boundary layer theory can be used to compute the condensate wave  
function close to the Thomas Fermi radius \cite{bdla}, but we shall not need this  
improvement in the present paper.

\subsection{Quasi-particle excitations}
Let us now turn to a Bose condensate with excitations and adopt a description  
in the Heisenberg picture, where the field operator satisfies the equation of  
motion
\begin{equation}
 i\hbar\frac{\partial\hat\psi}{\partial  
t}=-\frac{\hbar^2}{2m}\nabla^2\hat\psi+U(\bbox{r})\hat\psi+
  \frac{4\pi\hbar^2 a}{m}\hat\psi^+\hat\psi\hat\psi\,.
\label{eq:2.4e}
\end{equation}
For a weakly interacting Bose condensed gas the Bogoliubov approximation
consists in splitting the total Bose field operator $\hat{\psi}(t)$ into
\begin{equation}
 \hat{\psi}(t)=(\psi_0+\hat{\phi}(t))e^{-i\mu t/\hbar}\,,
\label{eq:2.14}
\end{equation}
where $\psi$ is the time-independent condensate wave function and  
$\hat{\phi}(t)$ describes the
excitations from the condensate, and to linearize the field equation for
$\hat{\phi}(t)$
\begin{equation}
 i\hbar\frac{\partial\hat{\phi}}{\partial t}=
  \Bigg(-\frac{\hbar^2}{2m}\nabla^2
   + U(\bbox{r})-\mu+\frac{8\pi\hbar^2a}{m}|\psi_0(\bbox{r})|^2\Bigg)
    \hat{\phi}
   + \frac{4\pi\hbar^2a}{m}\psi_0(\bbox{r})^2\hat{\phi}^+\,.
\label{eq:2.15}
\end{equation}

The Bogoliubov transformation
\begin{equation}
 \hat{\phi}(t)= {\sum_j}'\bigg(u_j(\bbox{r})\alpha_je^{-i\omega_jt}-
  v_j^*(\bbox{r})\alpha_j^+e^{i\omega_jt}\bigg)
\label{eq:2.16}
\end{equation}
and its adjoint, with Bose operators $\alpha_j$, $\alpha_j^+$ for the
quasi-particle excitations and with the sum over $j$ restricted to states with
$\omega_j\ne 0$, yields the Bogoliubov equations
\begin{equation}
 \left(\begin{array}{cc}
         -\frac{\hbar^2}{2m}\nabla^2+U_{\rm eff}(\bbox{r})-E_j &
         -K(\bbox{r})\\
          -K^*(\bbox{r}) &
            -\frac{\hbar^2}{2m}\nabla^2+U_{\rm eff}(\bbox{r})+E_j
             \end{array}\right)
             {u_j \choose v_j}=0,
\label{eq:2.18}
\end{equation}
where we used the abbreviations
\begin{eqnarray}
 && E_j=\hbar\omega_j\quad,\quad U_{\rm eff}(\bbox{r})=U(\bbox{r})
  +\frac{8\pi\hbar^2a}{m}|\psi_0(\bbox{r})|^2-\mu\nonumber\\
 && K(\bbox{r})=\frac{4\pi\hbar^2a}{m}\psi_0(\bbox{r})^2\,.
 \label{eq:2.19}
\end{eqnarray}
Eq.~(\ref{eq:2.18}) is consistent with and its solutions must be chosen to
satisfy the orthonormality conditions
\begin{eqnarray}
 && \int d^3r(u_ju_k^*-v_jv_k^*)=\delta_{jk}\nonumber\\
 && \int d^3r(u_j^*v_k-u_k^*v_j)=0
\label{eq:2.20}
\end{eqnarray}
in order to guarantee the Bose commutation relations of the $\alpha_j$,
$\alpha_j^+$.

A formal solution of eq.~(\ref{eq:2.18}) at zero energy $E_j=0$ is given
by the condensate
\begin{equation}
 u_j=v_j^*=\psi_0(\bbox{r})\quad,\quad E_j=0\,,
\label{eq:2.21}
\end{equation}
but this solution is not normalizable in the required sense. For a harmonic
trap $U(\bbox{r})=\frac{1}{2}m\sum_{i=1}^3\omega_{0i}^2x_i^2$ three
linearly independent and correctly normalized exact solutions
\begin{equation}
 u_i=b_i^+\psi_0(\bbox{r})\quad,\quad v_i=b_i\psi_0^*(\bbox{r})\quad,\quad
  E_i=\hbar\omega_{0i}
\label{eq:2.22}
\end{equation}
always exist \cite{Hutchinson}, where
\begin{equation}
 b_i=\sqrt{\frac{\hbar}{2m\omega_{0i}}}\frac{\partial}{\partial x_i}
   +\sqrt{\frac{m\omega_{0i}}{2\hbar}}x_i\quad,\quad b_i^+=(b_i)^+
\label{eq:2.23}
\end{equation}
are the Bose operators for the harmonical oscillations  in the trap.

\section{WKB-approximation for quasi-particle states in the classically  
allowed region}
\label{sec:WKB}
We shall here treat the WKB approximation for the case of an isotropic  
harmonic trap, where $U_{\rm eff}(r)$ and $K(r)$ depend only on the radial  
coordinate $r$.

Let us drop the index $j$ for the time being and expand formally in $\hbar$
\begin{eqnarray}
 u &=& (a_0(r)+\hbar  
a_1(r)+\hbar^2a_2(r)+\dots)e^{iS(r)/\hbar}Y_{lm}(\theta,\phi)\nonumber\\
 v &=& (b_0(r)+\hbar b_1(r)+\hbar^2b_2(r)+\dots)e^{iS(r)/\hbar}Y_{lm}
 (\theta,\phi)\,.
\label{eq:3.1}
\end{eqnarray}
In Eq.~(\ref{eq:3.1}) we shall treat $U_{\rm eff}(r)$ and $K(r)$
as being independent of $\hbar$. The angular dependence is separated out by  
the spherical harmonics $Y_{lm}$. The angular momentum, denoted by $J$ in the  
following, is then quantized in the ususal way by $J^2=\hbar^2 l(l+1)$, which  
will be replaced in our WKB-treatment according to Langer's rule  
$J^2=\hbar^2(l+1/2)^2$. Inserting the ansatz (\ref{eq:3.1}) into 
(\ref{eq:2.18})
we obtain up to order $\hbar^2$ the equations
\begin{eqnarray}
 \label{eq:3.2}
\bbox{L}_0{a_0\choose b_0}&=&0\\
  \bbox{L}_0{a_1\choose b_1} &=&-\bbox{L}_1{a_0\choose b_0}\label{eq:3.3}\\
 \bbox{L}_0{a_2\choose b_2}&=&-\bbox{L}_1{a_1\choose b_1}-
  \bbox{L}_2{a_0\choose b_0}
\label{eq:3.4}
\end{eqnarray}
with
\begin{eqnarray}
 \label{eq:3.5}
\bbox{L}_0 &=& \left(\begin{array}{cc}
         \frac{p_r^2}{2m}+\frac{J^2}{2mr^2}+U_{\rm eff}(r)-E &
         -K(r)\\
          -K^{*}(r) &
             \frac{p_r^2}{2m}+\frac{J^2}{2mr^2}+U_{\rm eff}(r)+E
             \end{array}\right)\\
\bbox{L}_1 &=& -\frac{i}{m}
  \left(p_r\frac{d}{d r}+\frac{1}{2}
   \frac{d p_r}{d r}+\frac{p_r}{r}\right)
   \left(\begin{array}{cc}
         1 & 0\\
         0 & 1\end{array}\right)\label{eq:3.6}\\
\bbox{L}_2 &=& -\frac{1}{2m}(\frac{d^2}{d  
r^2}+\frac{2}{r}\frac{d}{d r})
    \left(\begin{array}{cc}
         1 & 0\\
         0 & 1\end{array}\right)\label{eq:3.7}
\end{eqnarray}
where $p_r=d S/d r$.

The solvability condition of (\ref{eq:3.2}), the vanishing of the determinant  
of $L_0$, gives the condition
\begin{equation}
 \det (\bbox{L}_0)=
  \left(\frac{1}{2m}\left(\frac{d S}{d r}\right)^2+
   \frac{J^2}{2mr^2}+U_{\rm eff}(r)\right)^2-E^2-|K(r)|^2=0\,.
\label{eq:3.8}
\end{equation}
We can solve this equation for $E$, restricting ourselves to the branch
$E>0$, because the replacement $E\to-E$ merely amounts to the renaming
$u\to v^*$, $v\to u^*$. Then we obtain the time-independent Hamilton-Jacobi
equation
\begin{equation}
 H(d S/d r,r)=E
\label{eq:3.9}
\end{equation}
with
\begin{equation}
H(p_r,r)=
   \sqrt{\left(\frac{p_r^2}{2m}+\frac{J^2}{2mr^2}+U_{\rm eff}(r)\right)^2-
   |K(r)|^2} \,.
\label{eq:3.10}
\end{equation}
This defines the Hamiltonian for the classical quasi-particle dynamics.
Introducing the time-dependent action
\begin{equation}
 W(r,t)=S(r)-Et
\label{eq:3.11}
\end{equation}
we may replace eq.~(\ref{eq:3.9}) by the time-dependent Hamilton Jacobi
equation
\begin{equation}
 \frac{\partial W}{\partial t}+H(\frac{\partial W}{\partial r},r)=0\,.
\label{eq:3.12}
\end{equation}
It follows from eq.~(\ref{eq:3.8}) that with $S$ also $-S$ is a solution
of the Hamilton Jacobi equation (\ref{eq:3.9}).
It is interesting to note that the classical quasi-particle dynamics is
integrable only in the special case of an isotropic harmonic trap. Even in
the experimentally realized case of an anisotropic axially symmetric trap
the classical quasi-particle dynamics turns out to be nonintegrable, in  
general. A detailed investigation is given in \cite{Fliesser2}.
In this case a semi-classical quantization
of the quasi-particle dynamics requires the methods introduced in the field
of quantum chaos, such as semiclassical quantization based on periodic
orbits, or even replacing the quasi-particle spectrum and wave-functions by
results based on random matrix theory.

After satisfying eq.~(\ref{eq:3.8}) the general solution of
eq.~(\ref{eq:3.2}) can be written in the form
\begin{equation}
 {a_0\choose b_0}=a_0(r)\left(\begin{array}{l}
 1\\ \frac{-E+\sqrt{E^2+|K(r)|^2}}{K(r)}\end{array}\right)
\label{eq:3.13}
\end{equation}
where $a_0(r)$ is still arbitrary.

$\bbox{L}_0$ is not invertible. The solvability condition of 
eq.~(\ref{eq:3.3}) is that its inhomogeneity is orthogonal to the kernel of $\bbox{L}_0$
\begin{equation}
 (a^*_0,b^*_0)\bbox{L}_1{a_0\choose b_0}=0\,.
\label{eq:3.14}
\end{equation}
Evaluating this condition for $a_0(r)\ne 0$ we obtain after some
rearrangements a classical transport equation for $a_0$,
\begin{equation}
 \frac{d S}{d r}\frac{d}{d r}\ln
 \left(|a_0(r)|^2\left(1+
 (\sqrt{1+E^2/|K(r)|^2}-E/|K(r)|)^2\right)\right)
+\frac{1}{r^2}\frac{d}{d r}\left( r^2\frac{d S}{d r}\right)=0,
\label{eq:3.15}
\end{equation}
which can be rewritten as a continuity equation
\begin{equation}
 \frac{1}{r^2}\frac{d}{d r}\left[ r^2|a_0(r)|^2\left(1+
  \left(\sqrt{1+E^2/|K(r)|^2}-E/|K(r)|\right)^2\right)
  \frac{d S}{d r}\right]=0\,.
\label{eq:3.16}
\end{equation}
The physical meaning of this result as classical conservation law for the
quasi-particle current becomes clear if we rewrite it once more as
\begin{equation}
 \frac{1}{r^2}\frac{d}{d r}\left[ r^2\left(|a_0(r)|^2-|b_0(r)|^2\right)
  v_r(r)\right]=0,
\label{eq:3.17}
\end{equation}
where we used the identities
\begin{equation}
v_r(r)\equiv
 \left(\frac{\partial}{\partial p_r}
H(p_r,r)\right)_{p_r=d S/d r}=
 \frac{\sqrt{E^2+|K(r)|^2}}{E}\frac{1}{m}\frac{d S}{d r}=
  \frac{1}{m}
   \frac{|K|^2+(\sqrt{E^2+|K|^2}-E)^2}{|K|^2-(\sqrt{E^2+|K|^2}-E)^2}
   \frac{d S}{d r}
\label{eq:3.18}
\end{equation}
to express $d S/d r$ in terms of the radial component $v_r(r)$
of the quasi-particle velocity field. The general solution of  
eq.~(\ref{eq:3.17}) can
therefore be written as
\begin{equation}
 |a_0|^2-|b_0|^2=\frac{const}{r^2 v_r(r)}
\label{eq:3.19}
\end{equation}

In principle, this procedure can be continued. Having satisfied the
solvability condition for eq.~(\ref{eq:3.3}) we can determine its general
solution in the form
\begin{equation}
 {a_1\choose b_1}=a_1(r)\left(\begin{array}{l}
 1\\ -E/|K|+\sqrt{1+E^2/|K|^2}\end{array}\right)+
  {0\choose b_1^{\rm part}(r)}\,.
\label{eq:3.21}
\end{equation}
Here the first term is the general solution of the homogeneous equation
whose still arbitrary amplitude $a_1(r)$ must be determined from
the solvability condition of eq.~(\ref{eq:3.4})
\begin{equation}
 \left(a^*_0,b^*_0\right)
 \left\{\bbox{L}_1{a_1\choose b_1}+\bbox{L}_0{a_0\choose b_0}\right\}=0\,.
\label{eq:3.22}
\end{equation}
The second term in eq.~(\ref{eq:3.21}) is a particular solution of
eq.~(\ref{eq:3.3}), which we choose with vanishing amplitude
$a_1^{\rm part}(r)=0$ in order to define the first term in
eq.~(\ref{eq:3.21}) in an unambiguous way. As we shall not need the explicit
form of $a_1$, $b_1$ we shall here not work it out from
eqs.~(\ref{eq:3.21}), (\ref{eq:3.22}).

\section{Semiclassical quasi-particle energies in isotropic harmonic traps}
\label{sec:bohr}
For the radial canonical momentum as a function of $r$ 
we obtain from (\ref{eq:3.10})
\begin{equation}
 p_r=\sqrt{2m}\sqrt{\sqrt{E^2+K^2(r)}-\frac{J^2}{2mr^2}-U_{\rm eff}(r)}\,.
\label{eq:4.2}
\end{equation}
The semi-classical quantization can now be performed in a straightforward
manner by putting for the angular momentum
\begin{equation}
 J=J_\ell=\hbar(\ell+1/2)
\label{eq:4.3}
\end{equation}
and for the radial action variable
\begin{equation}
 I_r(E,J)=\frac{1}{\pi}\int_{{\textstyle r}_<}^{{\textstyle r}_>}
  p_r(E,J,r)dr=\hbar(n_r+1/2)\,.
\label{eq:4.4}
\end{equation}
Here $\ell$, $n_r$ are the integer angular momentum and radial quantum
number, respectively. We used `Langer's rule' \cite{La} of semi-classical
angular momentum
quantization by including $1/2$ on the right hand side of (\ref{eq:4.3}),
and took into account a phase-shift of $\pi/4$ in the wave function at the
classical turning points $r_<$, $r_>$ of the radial motion by including
$1/2$ on the right hand side of (\ref{eq:4.4}).

For the Thomas Fermi approximation of the condensate
the radial integral in eq.~(\ref{eq:4.4}) can be performed analytically,
even though the resulting expressions become rather lengthy. In this
manner an implicit analytical formula for the energy levels in WKB
approximation is obtained. Numerically solving the implicit expression for
$E$ for different values of $n_r$, $\ell$ we obtain the energy levels
$E_{{n_r}l}$ in WKB approximation.

\subsection{Types of motion}
In Thomas-Fermi approximation (\ref{eq:2.4}) we have
\begin{eqnarray}
 U_{\rm eff}(r) &=& \left|\mu-\frac{m}{2}\omega_0^2r^2\right|\nonumber\\
 K^2(r)         &=& \left(\mu-\frac{m}{2}\omega_0^2r^2\right)^2
                     \Theta\left(\mu-\frac{m}{2}\omega_0^2r^2\right)
\label{eq:4.5}
\end{eqnarray}
where
\begin{eqnarray}
 \mu &=& \frac{\hbar\omega_0}{2}R_0^2\quad,\nonumber\\
 R_0 &=& \left(\frac{15{N_0a}}{d_0}\right)^{1/5}=\frac{r_0}{d_0},\quad
 d_0=\sqrt{\frac{\hbar}{m\omega_0}}\,.
\label{eq:4.6}
\end{eqnarray}
The Thomas-Fermi approximation requires $R_0^2>>1$ for consistency. Let
us introduce scaled variables and parameters by
\begin{equation}
 \rho=(r/d_0R_0)\quad,\quad \epsilon=E/\mu\quad,\quad j=(J/\hbar R_0^2)
   \quad,\quad\pi_r=p_r(d_0/\hbar R_0)\,.
\label{eq:4.7}
\end{equation}
Then eq.~(\ref{eq:4.2}) becomes
\begin{equation}
 \pi_r=\sqrt{\sqrt{\epsilon^2+(1-\rho^2)^2\Theta(1-\rho)}-j^2/\rho^2-
  |1-\rho^2|}\,.
\label{eq:4.8}
\end{equation}
In order to find the turning points $\rho_<$, $\rho_>$ we have to look for the
solutions of $\pi_r=0$.\\
i) Domain $\rho<1$: Turning points in this region have to satisfy
\begin{equation}
 \rho_<^2=\frac{j^2}{\epsilon^2+2j^2}\left(1+
  \sqrt{1+\epsilon^2+2j^2}\right)\,.
\label{eq:4.9}
\end{equation}
The negative root is dismissed because only real and 
positive solutions are physical.
The remaining solution (\ref{eq:4.9}) turns out to be a lower boundary of
the $\rho$-values for which $\pi_r$ is real. The condition
\begin{equation}
 \rho_<^2<1\qquad\mbox{(\rm type B motion)}
\label{eq:4.10}
\end{equation}
is only fulfilled if
\begin{equation}
 \epsilon>j^2\qquad\mbox{(\rm type B motion)}\,.
\label{eq:4.11}
\end{equation}
We shall denote this motion as being of type B. As there is only a lower
turning point $\rho_<$ within the region $\rho<1$, the upper turning point
must lie in the region $\rho>1$, i.e. for type B motion 
the quasi-particle leaves and reenters the condensate during
each radial period.\\
ii) Domain $\rho>1$: Turning points in this region must satisfy
\begin{equation}
 \left.\begin{array}{l}
 \rho_>\\
 \rho_<\end{array}\right\}=\frac{1}{2}(\epsilon +1)\pm\frac{1}{2}
   \sqrt{(\epsilon+1)^2-4j^2}\,.
\label{eq:4.12}
\end{equation}
They exist if and only if
\begin{equation}
 \epsilon\ge 2j-1
\label{eq:412a}
\end{equation}
where we recall that $\epsilon$ is restricted to $\epsilon>0$ by definition.
The lower turning point given by eq.~(\ref{eq:4.12}) exists only if it
satisfies
\begin{equation}
\rho_<^2>1\qquad\mbox{(\rm type A motion)}
\label{eq:4.13}
\end{equation}
in which case the classical quasi-particle dynamics is entirely confined to
the region outside the condensate. It is then simply the motion in the
harmonic potential of the trap. We call this motion being of type A. It
requires the conditions
\begin{equation}
 \epsilon>1\quad,\quad j^2>\epsilon\quad>\quad2j-1\quad
 \mbox{(\rm type A motion)}
\label{eq:4.14}
\end{equation}
all to be satisfied, see fig.~\ref{fig:1}.

If the turning point $\rho_<$ does not exist in the region $\rho>1$, then
the turning point $\rho_<$ must occur for $\rho<1$, i.e. the only alternative
of type A motion is motion of type B satisfying eqs.~(\ref{eq:4.10}),
(\ref{eq:4.11}). The upper turning point of both types of motion lies in the
region $\rho>1$ and is given by
\begin{equation}
 \rho_>^2=\frac{1}{2}(\epsilon +1)+\frac{1}{2}\sqrt{(\epsilon+1)^2-4j^2}
 \quad\mbox{(\rm type A, B motion)}\,.
\label{eq:4.15}
\end{equation}
It is easily checked that $\rho_>^2>1$ for both types of motion. 
In fig.~\ref{fig:1}
the parts of the angular momentum --- energy plane accessible to the two
types of motion are shown. Let us now
proceed to evaluate the radial action of both types of motion and quantize.

\subsection{Asymptotic limit of a large condensate $R_0\to\infty$}
The integral to be evaluated is
\begin{equation}
 I_r=\frac{\hbar R_0^2}{\pi}\int_{\rho_<}^{\rho_>}
 \pi_r d\rho\,.
\label{eq:4.16}
\end{equation}
In order to obtain simple closed expressions we first study the case of
a large condensate for which $R_0\to\infty$, while we keep the energy $E$
and angular momentum $J$ of the quasi-particle fixed, i.e. the scaled
quantities
\begin{equation}
 \epsilon=\frac{2E}{\hbar\omega_0}\,\frac{1}{R_0^2}\quad,\quad
  j=\frac{\ell+1/2}{R_0^2}
\label{eq:4.17}
\end{equation}
are both $0(R_0^{-2})$ in that limit. It follows that the necessary condition
for type A motion $\epsilon\le j^2$ (i.e. motion which remains outside the
condensate) cannot be fulfilled in this limit, and therefore only type B
motion can occur. In the present limit we have asymptotically
\begin{eqnarray}
\label{eq:4.18}
 \rho_< \simeq &&
  \frac{\hbar\omega_0(\ell+1/2)}{\sqrt{2E^2+(\hbar \omega_0)^2
  (\ell+1/2)^2}}\,,\quad
   \rho_> \simeq \sqrt{1+\epsilon}\\
    \pi_r \simeq && \Theta(1-\rho)\frac{1}{R_0^2}
     \sqrt{\frac{2E^2}{\hbar^2\omega_0^2}\frac{1}{(1-\rho^2)}-
      \frac{(\ell+1/2)^2}{\rho^2}}\label{eq:4.20}\\
       && + \Theta(\rho-1)\sqrt{\epsilon+1-\rho^2}\,.\nonumber
\end{eqnarray}
The integral
\begin{equation}
 I_>=\frac{\hbar R_0^2}{\pi}\int_1^{\sqrt{1+\epsilon}}\sqrt{1+\epsilon-\rho^2}
  d\rho
\label{eq:4.21}
\end{equation}
can be performed asymptotically for $\epsilon\to 0$ with the result
\begin{equation}
I_>=\frac{\hbar}{3\pi}\,\frac{1}{R_0}
 \left(\frac{2E}{\hbar\omega_0}\right)^{3/2}\to 0\,.
\label{eq:4.22}
\end{equation}
The contribution of the part of the type B motion outside the condensate to
the total radial action therefore becomes negligible $0(R_0^{-1})$ as
$R_0\to\infty$. The 0(1)-part of the action is therefore given by
\begin{equation}
 I_r\simeq\frac{\hbar}{\pi}\int_{\textstyle{\rho}_<}^1
  \sqrt{\frac{2E^2}{\hbar^2\omega_0^2}\,
   \frac{1}{(1-\rho^2)}-\frac{\left(\ell+\frac{1}{2}\right)^2}{\rho^2}}
    d\rho\,.
\label{eq:4.23}
\end{equation}
The substitution
\begin{equation}
 \rho^2=1-\frac{E^2}{E^2+(\ell+1/2)^2(\hbar\omega_0)^2/2}\sin^2\varphi
\label{eq:4.24}
\end{equation}
reduces the integral to a rational function of $\sin\varphi$ and we obtain
after integration
\begin{equation}
 I_r=\frac{\hbar}{2}\left(
  \sqrt{2(E/\hbar\omega_0)^2+(\ell+1/2)^2}-(\ell+1/2)\right)+
   0\left(\frac{1}{R_0}\right)\,.
\label{eq:4.25}
\end{equation}
From the semi-classical quantization rule $I_r=\hbar(n_r+1/2)$ we obtain the
energy levels
\begin{equation}
 E_{{n_r}\ell}=\hbar\omega_0(2n_r^2+2n_r\ell+3n_r+\ell+1)^{1/2}
  \left(1+0\left(\frac{1}{R_0}\right)\right)\,.
\label{eq:4.26}
\end{equation}
This result is remarkably close to the hydrodynamic result
\begin{equation}
E_{{n_r}\ell}^{{\rm hyd}}=\hbar\omega_0(2n_r^2+2n_r\ell+3n_r+\ell)^{1/2}
\label{eq:4.27}
\end{equation}
derived by Stringari \cite{6b} for the low lying levels 
in the hydrodynamic regime.
In that regime the WKB-method is not applicable, but for energies
$E_{{n_r}\ell}>>\hbar\omega_0$ the agreement between (\ref{eq:4.26}) and
(\ref{eq:4.27}) becomes very good. Thus we find that there is a common regime
of applicability of both the hydrodynamic approximation and the WKB
approximation, where the energy is sufficiently large to apply the WKB
approximation but not too large to invalidate the hydrodynamic approach.

\subsection{Large-energy limit}

For type A motion the classical particle never enters the condensate
and the Bohr-quantization (\ref{eq:4.4}) leads to 
\begin{equation}
 E_{{n_r}\ell}=\hbar\omega_0(2n_r+\ell+3/2)-\mu \equiv E_{n}^{osc}.
\end{equation}
which depends only on the pricipal quantum number $n=2n_r+\ell$ (see equation  (\ref{eq:4.28a}) below).
However, for type B motion, i.e., $\epsilon \ge j^2$,
$E_{n_r \ell}$ differs from $E_{n_r \ell}^{osc}$
\begin{equation}
E_{n_r \ell}=E_{n}^{osc}+\hbar \omega_0\delta_{n \ell}.
\label{eq:shift}
\end{equation}
The shift $\delta_{n \ell}$ lifting the degenerecy of the free oscillator levels is expected to be small in the
large-energy limit $E_{n_r \ell} \gg \mu$ and can be calculated 
in the following way:

For type B motion (\ref{eq:4.4}) can be written as
\begin{eqnarray}
n_r&+&1/2={R_0^2 \over \pi} \int_{\rho_<}^{\rho_>} \pi_r (\rho) \, d\rho =
{R_0^2 \over \pi} \int_{\rho_<}^1 \, d\rho 
\sqrt{\sqrt{\epsilon^2+(1-\rho^2)^2}-j^2/\rho^2-(1-\rho^2)} \nonumber\\
&&+{R_0^2 \over \pi}\int_{\rho_<^{(o)}}^{\rho_>} \, d\rho
\sqrt{\epsilon^2-j^2/\rho^2+(1-\rho^2)}-{R_0^2 \over \pi}
\int_{\rho_<^{(o)}}^1 \, d\rho \sqrt{\epsilon^2-j^2/\rho^2+(1-\rho^2)},
\label{eq:radint}
\end{eqnarray}
where $\rho_<^{(o)}$  and $\rho_>$ are the turning points for
$\pi(\rho)=\sqrt{\epsilon^2-j^2/\rho^2+(1-\rho^2)}$, i.e.,
for the free oscillator:
\begin{equation}
\rho_> \, , \rho_<^{(o)} = \sqrt{\left({\epsilon+1 \over 2} \right) \pm
\sqrt{\left({\epsilon +1 \over 2} \right)^2 -j^2}}.
\label{eq:turnp}
\end{equation}
The second integral in (\ref{eq:radint}) is trivial and 
using (\ref{eq:shift}) eq. (\ref{eq:radint}) can be written as
\begin{eqnarray}
\delta_{n}(\ell)&=&{2 R_0^2 \over \pi}\int_{\rho_<}^1 \, d\rho
\sqrt{\sqrt{\epsilon^2+(1-\rho^2)^2}-j^2/\rho^2-(1-\rho^2)} 
\nonumber\\
&&-{2 R_0^2 \over \pi} \int_{\rho_<^{(o)}}^1 \, d\rho
\sqrt{\epsilon^2-j^2/\rho^2+(1-\rho^2)}.
\label{eq:shift1}
\end{eqnarray}
For type B motion $j^2 \le \epsilon$, thus in the present
limit $\epsilon \to \infty$ and $a \equiv j^2/\epsilon$ is
kept fixed ($0 \le a \le 1$). Our small parameter
is $1/\epsilon$.
In this limit the two integrals in leading order
cancel each other. With the replacement $\rho^2=1-z$ the
next to leading approximation gives
\begin{equation}
\delta_{n}(\ell)= {R_0^2 \over \pi \sqrt{\epsilon}}
\int_0^{1-a} dz {z \over \sqrt{1-z-a}} =
{4 R_0^2 \over 3 \pi\sqrt{\epsilon}}(1-a)^{3/2}.
\label{eq:shift2}
\end{equation}
The shift is clearly tending to zero in the large energy
limit, thus one can replace the unscaled energy $E_{n_r \ell}$
with $E_{n}^{osc}$ on the right hand side of eq. 
(\ref{eq:shift2}):
\begin{equation}
\delta_{n}(\ell)={1 \over 3 \pi }
{\left[ (4 \mu / \hbar\omega_0)(n +3/2 -\mu/\hbar\omega_0)-
(\ell+1/2)^{2})\right]^{3/2} \over \left[ 
n+3/2 -\mu/\hbar\omega_0\right]^2},
\label{eq:shlel}
\end{equation} 
which is the asymptotics of the shift in the large energy
limit. This is the result presented in \cite{Csordas2}. For large values of $n=2n_r+2\ell$ it agrees with the perturbative result of \cite{Csordas1}.

\subsection{Radial action in the case of a finite condensate}
We now evaluate the radial action for finite $R_0$. The two types of motion
must be considered separately.\\
{\bf Type A motion}:\\
This type of motion requires $1\le2j-1\le\epsilon\le j^2$. Its radial action
is that of the free harmonic trap and easily obtained as
\begin{equation}
 I_r(\epsilon,j)=\frac{\hbar R_0^2}{\pi}
  \int_{{\textstyle \rho}_<}^{{\textstyle \rho}_>}\frac{d\rho}{\rho}
   \sqrt{-\rho^4+\rho^2(\epsilon+1)-j^2}=
    \frac{\hbar R_0^2}{2}\left(\frac{\epsilon+1}{2}-j\right)
\label{eq:4.28}
\end{equation}
where $\rho_>$, $\rho_<$ are given by the two roots (\ref{eq:4.12}).
Quantizing according to eqs.~(\ref{eq:4.3}), (\ref{eq:4.4}) and solving
for the energy levels we obtain
\begin{equation}
 E_{{n_r}\ell}=\hbar\omega_0(2n_r+\ell+3/2)-\mu
\label{eq:4.28a}
\end{equation}
i.e. the result which is obtained for the free trap, shifted by the change
of the ground state energy due to the condensate.\\
{\bf Type B motion}:\\
For this type of motion we have $j^2<\epsilon$. The radial action is given
by the integral
\begin{equation}
 I_r(\epsilon,j)=\frac{\hbar R_0^2}{\pi}
  \int_{{\textstyle \rho}_<}^{{\textstyle \rho}_>}d\rho
   \left(\left(\epsilon^2+(1-\rho^2)^2\Theta(1-\rho)\right)^{1/2}
   -\frac{j^2}{\rho^2}-|1-\rho^2|\right)^{1/2}
\label{eq:4.29}
\end{equation}
where $\rho_<$, $\rho_>$ are given by eqs.~(\ref{eq:4.9}) and (\ref{eq:4.15}),
respectively.

The integral can be performed in terms of elementary functions. We give
a summary of the necessary steps in the appendix. The result is most
succinctly expressed in terms of the functions
\begin{equation}
 \int_A^{\pi/2}\frac{\cos^2\varphi}{a+b\sin\varphi}d\varphi=
  \left\{\begin{array}{ll}
          J_>(A,a,b) & \qquad a^2>b^2\\
          J_<(A,a,b) & \qquad a^2<b^2\end{array}\right.
\label{eq:4.30}
\end{equation}

\begin{equation}
 \int_A^{\pi/2}\frac{\cos^2\varphi}{(a+b\sin\varphi)^2}d\varphi=
  \left\{\begin{array}{ll}
          K_>(A,a,b) & \qquad a^2>b^2\\
          K_<(A,a,b) & \qquad a^2<b^2\end{array}\right.
\label{eq:4.31}
\end{equation}
which can be written and programmed explicitely in terms of elementary
functions. The radial action is then obtained in the form
\begin{eqnarray}
 I_r(\epsilon,j)=\frac{\hbar R_0^2}{\pi}
 \Bigg\{&&\sum_{i=1}^3C_iJ_>(A_1,a_i,b_1)+C_4J_<(A_1,a_4,b_1)\nonumber \\
   &&+C_5K_<(A_1,a_4,b_1)+C_6J_>(A_2,a_5,b_2)\Bigg\}
 \label{eq:4.32}
\end{eqnarray}
with the parameters
\begin{eqnarray}
\label{eq:4.33}
 A_1 &=& \arcsin
  \left(\frac{\epsilon^2+j^2}{\epsilon\sqrt{1+\epsilon^2+2j^2}}\right)\\
A_2 &=& \arcsin
   \left(\frac{1-\epsilon}{\sqrt{(\epsilon+1)^2-4j^2}}\right)
    \label{eq:4.34}\\
b_1 &=& \epsilon\sqrt{1+\epsilon^2+2j^2}\label{eq:4.35}\\
b_2 &=& \sqrt{\left(\frac{\epsilon+1}{2}\right)^2-j^2}\label{eq:4.36}\\
a_{1,2} &=& \mp(\epsilon+j^2)\sqrt{\epsilon^2+1}+j^2(\epsilon-1)
   \label{eq:4.37}\\
a_3 &=& -\epsilon-\epsilon^2-2j^2\label{eq:4.38}\\
a_4 &=& \epsilon-\epsilon^2\label{eq:4.39}\\
a_5 &=& \frac{\epsilon+1}{2}\label{eq:4.40}
\end{eqnarray}
and the coefficients
\begin{eqnarray}
\label{eq:4.41}
 C_{1,2} &=& \frac{1}{4}\left(1-\epsilon\mp\sqrt{\epsilon^2+1}\right)
  \frac{\epsilon(1+\epsilon^2+2j^2)}{\sqrt{\epsilon+j^2}}\\
 C_3 &=& \frac{1}{4}
 \frac{\epsilon^2(1+\epsilon^2+2j^2)}{\sqrt{\epsilon+j^2}}
\label{eq:4.42}\\
C_4 &=& \frac{1}{4}(\epsilon-2)
 \frac{\epsilon(1+\epsilon^2+2j^2)}{\sqrt{\epsilon+j^2}}\label{eq:4.43}\\
 C_5 &=& \frac{1}{2}\sqrt{\epsilon+j^2}\epsilon^2(1+\epsilon^2+2j^2)
 \label{eq:4.44}\\
 C_6 &=& \frac{1}{8}(\epsilon+1)^2-\frac{1}{2}j^2\,.\label{eq:4.45}
\end{eqnarray}
The semiclassical approximations for the quasi-particle excitation energies
are now obtained by solving the equation
\begin{equation}
 I_r\left(\frac{2E}{\hbar\omega_0R_0^2},\frac{\ell+1/2}{R_0^2}\right)=
  \hbar(n_r+1/2)\,.
\label{eq:4.46}
\end{equation}
for $E$. This last step cannot be done analytically, but it is easily
accomplished numerically.

\section{Semiclassical wave functions}
\label{sec:waves}
For purely radial dynamics eq.~(\ref{eq:3.16}) reduces to
\begin{equation}
\frac{1}{r^2}\,\frac{d}{d r}
 \left[a_0^2(r)r^2\left(1+(\sqrt{1+E^2/K^2}-E/K)^2\right)p_r\right]=0
\label{eq:5.1}
\end{equation}
which is easily solved, putting $[\dots]=$ const in eq.~(\ref{eq:5.1}).
Inserting the result in eq.~(\ref{eq:3.13}) we obtain the prefactors of the
semiclassical wave function to lowest order. The exponential factors
$\exp(\pm iS/\hbar)$ in
(\ref{eq:3.1}) are combined to satisfy the usual matching conditions on
the lower turning point. After some rearrangement in the prefactors we
obtain
\begin{eqnarray}\left(\begin{array}{l}
  u_{{n_r}lm}(\bbox{r})\\
  v_{{n_r}lm}(\bbox{r})\end{array}\right)=\frac{C_0}{2r}&&
 \left(\begin{array}{l}\sqrt{\sqrt{E^2+K^2}+K}+\sqrt{\sqrt{E^2+K^2}-K}\\
 \sqrt{\sqrt{E^2+K^2}+K}-\sqrt{\sqrt{E^2+K^2}-K}\end{array}\right)
  \cdot
  \nonumber\\
   &&\qquad \frac{\sin\left(\frac{1}{\hbar}\int_{r_<}^rp_r(r)dr+
   \frac{\pi}{4}\right)}{\sqrt[4]{E^2+K^2}\sqrt{|p_r(r)|}}
    Y_{\ell m}(\vartheta,\varphi)
   \label{eq:5.2}
\end{eqnarray}
where $p_r(r)$, $K^2(r)$, are defined in eqs.~(\ref{eq:4.2}),
(\ref{eq:4.5}), and $r_<=\rho_<d_0R_0$ given by eq.~(\ref{eq:4.9}) for type
B motion or (\ref{eq:4.12}) for type A motion. The normalization condition
(\ref{eq:2.20}) serves to determine the constant $C_0$ by the integral
\begin{equation}
 1=\frac{C_0^2}{2}\int_{r_<}^{r_>}dr\frac{E}{\sqrt{E^2+K^2}}\,
  \frac{1}{|p_r(r)|}=\frac{C_0^2}{4m}T_r
\label{eq:5.3}
\end{equation}
where $T_r$ is the period of the radial motion, and $r_>=\rho_>d_0R_0$ is
defined by eq.~(\ref{eq:4.15}). To obtain eq.~(\ref{eq:5.3}) we have
restricted the normalization integral to the classically allowed domain
and replaced the rapidly oscillating $\sin^2(\hbar^{-1}\int^rp_rdr+\pi/4)$
by its average 1/2. Furthermore we used the identities (\ref{eq:3.18}) for
the radial velocity.

\subsection{General case}
The integral in eq.~(\ref{eq:5.3}) can be carried out explicitely. The
necessary steps are summarized in Appendix B. We obtain
\begin{equation}
 C_0^2=\left\{\begin{array}{l@{\qquad}l}
  \frac{4m\omega_0}{\pi} & \mbox{\rm type A motion}\\
   \frac{4m\omega_0}{\frac{\pi}{2}-\alpha+
      4\epsilon\beta(2\epsilon^2+4j^2)^{-1/2}}& \mbox{\rm type B motion}
   \end{array}\right.
\label{eq:5.4}
\end{equation}
with
\begin{eqnarray}
\alpha &=& \arctan\frac{1-\epsilon}{2\sqrt{\epsilon-j^2}}\nonumber\\
\beta  &=& \arctan
 \left[\frac{\sqrt{\epsilon-j^2}}{\sqrt{2(\epsilon^2+2j^2)}}\,
       \frac{\epsilon+\epsilon^2+2j^2+\epsilon\sqrt{1+\epsilon^2+2j^2}}
       {\epsilon^2+j^2+\epsilon\sqrt{1+\epsilon^2+2j^2}}\right]\,.
\label{eq:5.5}
\end{eqnarray}
Using these results it can be shown that for type B-motion the
wave-functions for large energy tend to those of the free harmonic trap.

\subsection{Large condensate}
The result (\ref{eq:5.4}) simplifies if we consider the limit $R_0\to\infty$,
keeping
\begin{equation}
 E=\frac{1}{2}\hbar\omega_0R_0^2\epsilon\,,\quad\ell+1/2=R_0^2j
\label{eq:5.6}
\end{equation}
fixed. Type A motion disappears in that limit and for type B motion we obtain
\begin{equation}
 C_0^2=\frac{2m\omega_0}{\pi}\sqrt{2+\hbar^2\omega_0^2(\ell+1/2)^2/E^2}\,.
\label{eq:5.7}
\end{equation}
The correctly normalized wave functions (\ref{eq:5.2}) can now be used to
evaluate matrix elements in the semiclassical limit. This will be done
in the next section to calculate the cross section for inelastic light
scattering by which a quasi-particle is created from the condensate.

\section{Semiclassical structure function}
\label{sec:scattering}
To introduce the dynamical structure function let us first consider the differential cross-section for the off-resonant scattering of light from the trapped gas.
Light with incoming wave-vector and frequency $\bbox{q}_L,\omega_L$ is scattered into an outgoing field with $\bbox{q}'_L=\bbox{q}_L-\bbox{k}, \omega'_L=\omega_L-\omega$ with $\omega<<\omega_L$. The scattering angle $\theta$ between $\bbox{q}'_L$ and $\bbox{q}_L$ is related to $k$ by the usual kinematical relation $ck=2\omega_L sin(\theta/2)$ which implies the allowed interval $0<k<2\omega_L/c$. The permitted interval for the transferred frequency $\omega$ can be estimated from the corresponding relation for free atoms $0<\omega<(2\hbar\omega^2_L/m c^2)$. Let now the differential cross-section for the off-resonant light scattering from a single atom in its ground state be given by $\left (d\sigma/d\Omega\right )_{At}$. Then the spectral distribution of the differential cross-section for light scattering off the trapped atomic gas is given by \cite{Jav,Ruo}
\begin{equation}
 \frac{d\sigma}{d\Omega d(\hbar\omega )}=\left (\frac{d\sigma}{d\Omega}\right )_{At} S(\bbox{k},\omega)  \,.
\label{eq:6.0}
\end{equation}

The dynamical structure function $S(\bbox{k},\omega)$ is the Fourier transform
of the density --- density correlation function and defined as the thermal
expectation value
\begin{equation}
 S(\bbox{k},\omega)=\frac{1}{Z}\sum_{\mu,\mu'}e^{-\beta E_\mu}
  |\langle\mu'|\rho(\bbox{k})|\mu\rangle|^2\delta(\hbar\omega+E_\mu-E_{\mu'})
  \,.
\label{eq:6.1}
\end{equation}
Here $|\mu\rangle$ and $E_\mu$ are energy eigenstates and levels of the system,
and
\begin{equation}
 \rho(\bbox{k})=\int d^3r\,\psi^+(\bbox{r})e^{-i\bbox{k}\cdot\bbox{r}}
  \psi(\bbox{r})
\label{eq:6.2}
\end{equation}
is the operator of the density fluctuation with wave vector $\bbox{k}$.

In the present case $S(\bbox{k},\omega)$ assumes the form
\begin{eqnarray}
 S(\bbox{k},\omega)&=&S_0(\bbox{k})\delta(\hbar\omega)+
  {\sum_i}'S_i(\bbox{k})\left[\delta(\hbar\omega+E_i)+
   e^{-\beta E_i}\delta(\hbar\omega-E_i)\right]
   \nonumber \\
  &&+{\sum_{ij}}' S_{ij}^{(1)}(\bbox{k}) \left[
  \delta(\hbar\omega+E_i+E_j)+e^{-\beta(E_i+E_j)}
  \delta(\hbar\omega-E_i-E_j)\right]
   \nonumber \\
  &&+{\sum_{ij}}' S_{ij}^{(2)}(\bbox{k}) \Theta(E_i-E_j)\left[
   \delta(\hbar\omega+E_i-E_j)+e^{-\beta(E_i-E_j)}
   \delta(\hbar\omega-E_i+E_j)
  \right],
\label{eq:6.3}
\end{eqnarray}
where the sums $\sum'$ run over the quasi-particle states with $E_i\ne 0$. 
The first term in eq.~(\ref{eq:6.3}) describes
coherent elastic scattering, and $S_0(\bbox{k})$ is simply the square of the
Fourier transform of the density
\begin{equation}
 S_0(\bbox{k})=
  \left|\int d^3re^{-i\bbox{k}\cdot\bbox{r}}\left\{|\psi_0(\bbox{r})|^2+
   {\sum_i}'|v_i(\bbox{r})|^2
   +{\sum_i}' \bigl(|u_i|^2+|v_i|^2 \bigr) \bigl( 
   e^{\beta\hbar\omega_i}-1\bigr)^{-1}\right\}\right|^2\,.
\label{eq:6.4}
\end{equation}
Because  $\psi_0 \propto \sqrt{N_0}$ 
this is by far the dominant contribution in
$S(\bbox{k},\omega)$. This contribution of the scattered light, by its interference with the non-scattered light, is used in the phase-contrast
imaging of the condensates \cite{pc}. The part $\sum'_i|v_i(\bbox{r})|^2$ gives the number
density of particles outside the condensate at temperature $T=0$ due to the
interaction. It can be easily seen from eq.~(\ref{eq:5.2}) 
that $|v_i(\bbox{r})|^2$ vanishes for $r>r_0$. For weak
interaction this number density is much smaller than $N_0/r_0^3$. 

The
second term in eq.~(\ref{eq:6.3}) describes the
creation of a quasi-particle from the condensate. The matrix element for
this process is
\begin{equation}
 M_i^*(\bbox{k})=\int d^3r\, e^{i\bbox{k}\cdot\bbox{r}}\psi_0
  (\bbox{r})(u_i^*(\bbox{r})-v_i^*(\bbox{r}))
\label{eq:6.5}
\end{equation}
and
\begin{equation}
 S_i(\bbox{k})=|M_i(\bbox{k})|^2(\bar{n}_i +1), \quad 
\bar{n}_i={1\over e^{\beta E_i}-1}
\label{eq:6.6}
\end{equation}
The other quantities in (\ref{eq:6.3}) are:
\begin{eqnarray}
M_{ij}^{(1)*}(\bbox{k})= -\int d^3r\, u_i^* v_i^* e^{i\bbox{k}\cdot\bbox{r}}
&,&\quad M_{ij}^{(2)*}(\bbox{k})=\int d^3r\, (u_i^* u_j+v_i^* v_j) 
e^{i\bbox{k}\cdot\bbox{r}} \nonumber \\
S_{ij}^{(1)}(\bbox{k})=\left| M_{ij}^{(1)} (\bbox{k})\right|^2
\sqrt{(\bar{n}_i+1)(\bar{n}_j+1)}&,& \quad
S_{ij}^{(2)}(\bbox{k})=\left| M_{ij}^{(2)} (\bbox{k})\right|^2
\sqrt{(\bar{n}_i+1)\bar{n}_j}
\label{eq:6.6a}
\end{eqnarray}
Here $S_{ij}^{(1)}$ is proportional to the cross-section for the absorption of an energy $\hbar \omega$ by creating a pair of quasi-particles with energies $E_i, E_j$ from the condensate. The cross-section for the time-reversed process is smaller by a factor $e^{-\beta (E_i +E_j)}$ due to detailed balance.
Similarly $S_{ij}^{(2)}\Theta (E_i -E_j)$ describes the cross-section for absorbing an energy $\hbar \omega$ by converting a quasi-particle of energy $E_j$ into one with energy $E_i$. Both $S_{ij}^{(1)}$ and $S_{ij}^{(2)}$ give rise to a faint and broad spectral background in the scattering. In the following we shall therefore content ourselves with the evaluation of $|M_i(\bf k)|^2$.

\subsection{General case}
Here we shall evaluate the matrix element (\ref{eq:6.5}) in the semiclassical
limit. It is clear that only type B motion can contribute, because otherwise
$\psi_0$ and $(u_i-v_i)$ don't overlap. The condensate wave function is
given by eq.~(\ref{eq:2.4}). The amplitude $(u_i-v_i)$ is obtained from
eq.~(\ref{eq:5.2}), and $e^{-i\bbox{k}\cdot\bbox{r}}$, where $\bbox{k}$ in
$z$-direction in semiclassical approximation is expanded as
\begin{eqnarray}
e^{-ikz} &=& \sum_{\ell=0}^\infty (-i)^\ell\frac{2}{r}
 \sqrt{\frac{\pi(2\ell+1)}{k}}
  \frac{1}{\sqrt[4]{k^2-\left(\frac{\ell+1/2}{r}\right)^2}}\nonumber\\
 &&\,\cos\left(\int_{\left(\frac{\ell+1/2}{k}\right)}^r
  \sqrt{k^2-\left(\frac{\ell+1/2}{r}\right)^2}-\frac{\pi}{4}\right)
   Y_{\ell 0}(\Theta,\Phi)\,.
\label{eq:6.7}
\end{eqnarray}
The integral (\ref{eq:6.5}) is to be taken over the angles $\Theta,\Phi$
selecting the angular momentum quantum number from the sum in (\ref{eq:6.4})
and over $r^2dr$ from $r=r_<$ to $r=r_0$. This radial integral is of the
general form
\begin{equation}
 \int_{r_<}^{r_0}dr\,f(r)\cos\varphi(r)\simeq
  \sqrt{\frac{2\pi}{|\varphi''(z_0)|}}f(z_0)\cos
   \left(\varphi(z_0)+\mbox{\rm sgn}
    \left(\varphi''(z_0)\right)\frac{\pi}{4}\right)
\label{eq:6.8}
\end{equation}
where the evaluation is done in the stationary phase approximation and $z_0$
is defined as solution of $\varphi'(z_0)=0$ in the interval $r_<<z_0<r_0$.
Before writing the integral explicitely we introduce scaled parameters and
variables as before and in addition
\begin{equation}
 k=R_0\kappa/d_0\quad,\quad C_0=\sqrt{m\omega_0\tilde{C}}\,.
\label{eq:6.9}
\end{equation}
Then, neglecting a rapidly oscillating term without stationary phase in the
interval $r_<<r<r_0$, the amplitude can be written as
\begin{eqnarray}
  M_{{n_r}lm}(k)&=& \delta_{m0}(-i)^\ell\sqrt{\frac{\tilde{C}}{\kappa}}
  \frac{1}{R_0^2}\sqrt{\frac{15N_0(2\ell+1)}{8}}
   \int_{{\displaystyle \max(j/\kappa,\rho_<)}}^1d\rho
    \sqrt{1+\frac{1-\rho^2}{\sqrt{\epsilon^2+(1-\rho^2)^2}}}\nonumber\\
 &&  \frac{\sqrt{1-\rho^2}}{\sqrt{\pi_w(\rho)}}\cdot
      \frac{1}{\sqrt{\pi_r(\rho)}}\cos
    \left[R_0^2\int_{{\displaystyle \rho_<}}^{\displaystyle \rho}
     d\rho'\pi_r(\rho')-R_0^2
         \int_{j/\kappa}^\rho d\rho'\pi_w(\rho')\right]\,.
  \label{eq:6.10}
\end{eqnarray}
with $\pi_r(\rho)$ given by eq.~(\ref{eq:4.8}) and
\begin{equation}
 \pi_w(\rho) = \sqrt{\kappa^2-j^2/\rho^2}
\label{eq:6.11}
\end{equation}
The stationary point $z_0$ is determined by
\begin{equation}
 \pi_r(z_0)=\pi_w(z_0)
\label{eq:6.12}
\end{equation}
which is solved by
\begin{equation}
 z_0=\sqrt{\frac{\kappa^4+2\kappa^2-\epsilon^2}{2\kappa^2}}\,.
\label{eq:6.13}
\end{equation}
Now we have to examine under which conditions $z_0$ lies in the required
interval:\\
i) The condition for $\psi_0$, $u-v$, and $e^{-ikz}$ to overlap
are
\begin{equation}
 \epsilon>j^2\quad,\quad j<\kappa
\label{eq:6.14}
\end{equation}
ii) $z_0$ satisfies $0\le z_0^2\le 1$ if
\begin{equation}
 \kappa^4+2\kappa^2\ge\epsilon^2\ge\kappa^4
\label{eq:6.15}
\end{equation}
iii) $z_0$ satisfies the lower bound $j/\kappa\le z_0$ if
\begin{equation}
 \epsilon^2\le\kappa^4+2\kappa^2-2j^2
\label{eq:6.16}
\end{equation}
which is stronger than the first inequality (\ref{eq:6.15}) if the second
inequality (\ref{eq:6.15}) holds.\\
iv) $z_0$ satisfies the lower bound $\rho_<\le z_0$ if
\begin{equation}
 \sqrt{1+\epsilon^2+2j^2}\le 1+
 \frac{\epsilon^2(\kappa^4+2\kappa^2-2j^2-\epsilon^2)+2j^2\kappa^4}
      {2j^2\kappa^2}\,.
\label{eq:6.17}
\end{equation}
It follows from (\ref{eq:6.16}) that the right hand side of (\ref{eq:6.17})
is positive. Therefore both sides of (\ref{eq:6.17}) can be squared without
changing the inequality, and we obtain  after some rearrangements instead
of (\ref{eq:6.17})
\begin{equation}
 \left[j^2+
  \left(\frac{\epsilon^2-\kappa^4}{2\kappa^2}\right)\,
   \left(1-\frac{\epsilon^2-\kappa^4}{2\kappa^2}\right)\right]^2
 \le
 \left[
  \left(1-\frac{\epsilon^2-\kappa^4}{2\kappa^2}\right)\,
   \left(\frac{\epsilon^2+\kappa^4}{2\kappa^2}\right)\right]^2\,.
\label{eq:6.18}
\end{equation}
The terms inside the brackets $[\dots]$ on both sides of (\ref{eq:6.18})
are positive, by virtue of eq.~(\ref{eq:6.15}). Therefore we may take the
square root of both sides of eq.~(\ref{eq:6.18}), to obtain after some more
rearrangements, again the inequality (\ref{eq:6.16}). Thus,
eq.~(\ref{eq:6.17}) leads to no additional condition if inequalities
(\ref{eq:6.14})--(\ref{eq:6.16})
are satisfied, which are therefore 
the semiclassical selection rules for the creation
of quasi-particles from the condensate. For later reference it is usefull to rewrite the most restrictive of these inequalities in dimensional quantities as follows
\begin{equation}
 \frac{\hbar^2 k^2}{2 m} \le E_{n_r \ell} \le
 \sqrt{\left(\frac{\hbar^2 k^2}{2 m} \right)^2 +\mu \frac{\hbar^2 k^2}{m} - \frac{1}{2} (\ell +\frac{1}{2})^2(\hbar \omega_0)^2}\,.
\label{eq:6.20a}
\end{equation}
A numerical example for this selsction rule is shown in Fig.~\ref{fig:5}.

Next we need to determine $\pi_w(z_0)=\pi_r(z_0)$ and obtain
\begin{equation}
 \pi_w(z_0)=\kappa
  \sqrt{\frac{\kappa^4+2\kappa^2-\epsilon^2-2j^2}
   {\kappa^4+2\kappa^2-\epsilon^2}}\,.
\label{eq:6.19}
\end{equation}
We also need the sign of
\begin{equation}
 \varphi^{''}(z_0)=R_0^2(\pi'_r(z_0)-\pi'_w(z_0))=
 R_0^2 {2\kappa^2 \over \epsilon^2+\kappa^4}{\kappa^4+2\kappa^2-\epsilon^2
 \over \sqrt{2(\kappa^4+2\kappa^2-\epsilon^2-2j^2)}}
\label{eq:6.20}
\end{equation}
which is evaluated to be positive. We can now apply the formula
(\ref{eq:6.7}) to (\ref{eq:6.9}) and obtain
\begin{equation}
 M_{{n_r}lm}(k)=(-i)^\ell \delta_{m0}\sqrt{\frac{\tilde{C}}{\kappa}}\,
  \frac{\epsilon}{\kappa^3 R_0^3}\sqrt{\frac{15\pi N_0(2\ell+1)\sqrt{2}}{8}}\,
   \frac{\sqrt{\epsilon^2-\kappa^4}}
    {\sqrt[4]{\kappa^4+2\kappa^2-\epsilon^2-2j^2}}
     \cos\left(\varphi+\frac{\pi}{4}\right)
\label{eq:6.21}
\end{equation}
where the phase $\varphi$ 
\begin{equation}
\varphi=\varphi_1+\varphi_2 
\label{eq:6.21a}
\end{equation}
is given by the integrals
\begin{eqnarray}
 \varphi_1 &=& R_0^2\int_{\rho_<}^{z_0}d\rho\pi_r(\rho)
 \nonumber\\
 \varphi_2 &=& -R_0^2\int_{j/\kappa}^{z_0}d\rho\pi_w(\rho)\,.
\label{eq:6.22}
\end{eqnarray}
Integrals of this kind had already to be evaluated when calculating the
action $I_r$, and the same substitutions as were made there are also of help
here. In fact the integrals (\ref{eq:6.22}) and (\ref{eq:4.29}) differ
only by a prefactor $\hbar/\pi$ and by the upper boundary. Using this
observation we obtain by the series of steps summarized for
eq.~(\ref{eq:4.29}) in appendix A a result for (\ref{eq:6.22}) which
differs from (\ref{eq:a5}) only by the prefactor and the boundary
$A_1\to A_3$ with
\begin{equation}
 A_3=\arcsin
  \left[\frac{\epsilon-\kappa^2+\epsilon^2+\epsilon\kappa^2+2j^2}
   {(\epsilon+\kappa^2)\sqrt{1+\epsilon^2+2j^2}}\right]\,.
\label{eq:6.23}
\end{equation}
Explicitely
\begin{eqnarray}
 \varphi_1=R_0^2\bigg\{&&\sum_{i=1}^3C_iJ_>(A_3,a_i,b_1)+C_4J_<
           (A_3,a_4,b_1)\nonumber\\
           &&\quad +C_5K_<(A_3,a_4,b_1)\bigg\}
\label{eq:6.24}
\end{eqnarray}
with the parameters given by eqs.~(\ref{eq:4.35})--(\ref{eq:4.44}). The
second integral (\ref{eq:6.22}) is simplified by the substitution
\begin{equation}
 \rho=\frac{j}{\kappa\cos u}
\label{eq:6.25}
\end{equation}
and gives, after evaluation
\begin{equation}
 \varphi_2=-R_0^2j
  \left\{\sqrt{\frac{\kappa^4+2\kappa^2-\epsilon^2}{2j^2}-1}-\arccos
   \left(\frac{\sqrt{2}j}{\sqrt{\kappa^4+2\kappa^2-\epsilon^2}}
    \right)\right\}\,.
\label{eq:6.26}
\end{equation}
Now the matrix element is completely determined in the semiclassical
approximation. The inelastic part of the structure function (\ref{eq:6.3})
at $T=0$ can be given by the sum
\begin{eqnarray}
 S^{(1)}(\bbox{k},\omega)= {\pi d_0 \omega^2\over \sqrt{2} a (kd_0)^7 \hbar \omega_0^3}
  {\sum_{n_r}}'{\sum_\ell}' &&
   \frac{\tilde{C}_{{n_r}\ell}(2\ell+1)\left[(2
    \omega/\omega_0)^2-(kd_0)^4\right]}
     {\sqrt{(kd_0)^4+2(kd_0)^2R_0^2-\left(2
    \omega/\omega_0\right)^2-2
     \left(\ell+\frac{1}{2}\right)^2}}\nonumber\\
     && \cdot\left(\frac{1-\sin 2\varphi_{{n_r}\ell}}{2}\right)
      \delta\left(\frac{\omega+\omega_{{n_r}l}}{\omega_0}\right)
\label{eq:6.27}
\end{eqnarray}
where the two sums are restricted to quantum numbers $n_r$, $\ell$ satisfying
the semiclassical selection rules (\ref{eq:6.14})--(\ref{eq:6.16}) with $\epsilon=
\hbar\omega_{{n_r}\ell}/\mu$ and $j=(\ell+\frac{1}{2})/R_0^2$. Here
we have put indices $n_r$, $\ell$ on $\tilde{C}$ and $\varphi$ defined by eqs.(\ref{eq:6.9}),(\ref{eq:5.4}) and (\ref{eq:6.21a}), respectively, to indicate
the dependence of these quantities on the quantum numbers $n_r$,$\ell$ via $\epsilon=
\hbar\omega_{{n_r}\ell}/\mu$ and $j=(\ell+\frac{1}{2})/R_0^2$.

\subsection{Limit of large condensate}
In the limit of a large condensate $R_0>>1$, $E/\hbar\omega_0\sim0(1)$,
$(\ell+1/2)\sim0(1)$, $kR_0d_0\sim 0(1)$ the result (\ref{eq:6.27}) for the
inelastic structure function simplifies. For the prefactor
$\tilde{C}_{{n_r}\ell}$ we can use the asymptotic form (\ref{eq:5.7}) with
eq.~(\ref{eq:6.9}). The semiclassical selection rules
(\ref{eq:6.14})--(\ref{eq:6.16}) take the form
\begin{equation}
 E_{{n_r}\ell}\ge\frac{1}{2}\hbar\omega_0\frac{(\ell+1/2)^2}{R_0^2}\quad;
  \quad \ell+1/2\le R_0kd_0
\label{eq:6.28}
\end{equation}
\begin{equation}
 \frac{1}{2}\hbar\omega_0(kd_0)^2\le E_{{n_r}\ell}\le \frac{1}{\sqrt{2}}
  \hbar\omega_0\sqrt{R_0^2(kd_0)^2-(\ell+1/2)^2}
\label{eq:6.29}
\end{equation}
where the right-hand side of (\ref{eq:6.29}) has been simplified using
$(kd_0/R_0)^2<<1$.

The first condition of (\ref{eq:6.28}) is always satisfied for sufficiently
large $R_0$, the second represents a restriction on the size of $\ell$
since we want to assume $R_0kd_0\sim 0(1)$. In (\ref{eq:6.29}) we can use
the result (\ref{eq:4.26}) for the quasi-particle energies in large condensates
and obtain
\begin{equation}
 2n_r+ \ell+3/2\le R_0kd_0
\label{eq:6.30}
\end{equation}
for the second inequality in (\ref{eq:6.29}), while the first is always
satisfied in the limit we consider. It is somewhat surprising to see the
energy levels of the free trap appearing in (\ref{eq:6.30}). Under the
square-root in the denominator of eq.~(\ref{eq:6.27}) and also in the
nominator we may again neglect $(kd_0)^4$ compared to the other terms. Finally,
the tedious general results (\ref{eq:6.24}), (\ref{eq:6.26}) with
(\ref{eq:4.33})--(\ref{eq:4.44}) for $\varphi_1$ and $\varphi_2$ may be
simplified considerably by using the asymptotic results (\ref{eq:4.18}),
(\ref{eq:4.20}) for $\rho_<$ and $\pi_r$, and
\begin{equation}
z_0 \to \sqrt{1-2(\omega_{n,\ell}/\omega_0R_0kd_0)^2}
\label{eq:6.31}
\end{equation}

\begin{equation}
 \pi_w(\rho)=\frac{1}{R_0^2}\sqrt{(R_0kd_0)^2-(\ell+1/2)^2/\rho^2}
\label{6.32}
\end{equation}
following from eqs.~(\ref{eq:6.13}),  (\ref{eq:6.11}) in
eqs.~(\ref{eq:6.22}). The integrals are now much easier to
perform and we obtain after some further rearrangements
\begin{equation}
 \varphi_{{n_r}\ell}(k)=-R_0kd_0\left(\sin\chi_{{n_r}\ell}(k)-
  \chi_{{n_r}\ell}(k)\cos\chi_{{n_r}\ell}(k)\right)
\label{eq:6.33}
\end{equation}
with the angle $\chi_{{n_r}\ell}(k)$ defined by
\begin{equation}
 \cos\chi_{{n_r}\ell}(k)=\frac{2n_r+\ell+3/2}{R_0kd_0}\,.
\label{eq:6.34}
\end{equation}
The inequality (\ref{eq:6.30}) is incorporated in this definition. After all
this the structure function (\ref{eq:6.27}), in the limit of large
condensates, takes the explicit form
\begin{eqnarray}
 S^{(1)}(\bbox{k},\omega)= \frac{d_0}{a\hbar\omega_0}\,
 \frac{4}{(kd_0)^7}|\frac{\omega}{\omega_0}|^3
 \mathop{\sum_{n_r}\sum_\ell}_{(2n_r+\ell+3/2\le R_0kd_0)}&&
  \quad \frac{(2n_r +\ell+3/2)(2\ell+1)}
 {\sqrt{(R_0kd_0)^2-(2n_r+\ell+3/2)^2}} \label{eq:6.35}\\
&& \cos^2\left(\frac{\pi}{4}+\varphi_{{n_r}\ell}(k)\right)
    \delta\left((\omega+\omega_{{n_r}\ell})/\omega_0\right)\nonumber
\end{eqnarray}
with $\varphi_{{n_r}\ell}$ and $\omega_{{n_r}\ell}$ given by
eqs.~(\ref{eq:6.33}) and (\ref{eq:4.26}) respectively. It is interesting to
note that $S^{(1)}(\bbox{k},\omega)$ in the present limit has resonances
for the {\it maximum} momentum transfer $R_0kd_0=2n_r+\ell+3/2$ by
which states with the principal quantum number $n=2n_r+\ell$ can be excited.
The angles $\chi_{{n_r}\ell}(k)$ and $\varphi_{{n_r}\ell}(k)$ vanish there.
Similar resonances appear also in the more general result (\ref{eq:6.27})
which diverges whenever the border of the semiclassical selection rule
(\ref{eq:6.16}) is approached.

\section{Some numerical examples}
\label{sec:numerics}
All the calculations, so far, have been analytical, but the final results
for the energy levels, wave functions and inelastic structure function are
not obtained in explicit form and must be evaluated numerically from the
implicit results we have derived. This can be done without difficulty. We
shall here consider some examples of such results:

\subsection{Energy levels:}
The energy levels are obtained by solving eq.~(\ref{eq:4.46}) for $E$. This
can be done numerically without difficulty and ambiguity, since
\begin{equation}
 \frac{\partial I_r}{\partial E}=\frac{T_r}{2\pi}>0
\label{eq:7.1}
\end{equation}
where $T_r$ is the period of the radial motion. 
In Fig.~\ref{fig:2}  we take $(\mu/\hbar\omega_0)=\frac{1}{2}R_0^2=4.27$ 
and plot the level shifts $\delta_n(\ell)$ defined as a function
of $\ell$ and the {\it principal} quantum number $n=\ell+2n_r$ by
\begin{equation}
 \hbar\omega_0\delta_n(\ell)=E_{{n_r}\ell}+\mu-\hbar\omega_0(n+3/2)\,.
\label{eq:7.2}
\end{equation}
As for type A motion the levels are the same as for the free trap, apart from
a shift in the zero-point energy from $3\hbar\omega_0/2$ to 
$3\hbar\omega_0/2-\mu$,
it is clear that (\ref{eq:7.2}) vanishes for this type of motion,
characterized by the simultaneous inequalities $1\le2j-1\le\epsilon\le j^2$,
which imply
\begin{equation}
 (E_{{n_r}\ell}+\mu)/\hbar\omega_0\ge(\ell+1/2)\ge
  (2E_{{n_r}\ell}/\hbar\omega_0)^{1/2}(2\mu/\hbar\omega_0)^{1/2}\,.
\label{eq:7.3}
\end{equation}
The first inequality must always be satisfied for energy levels to exist, at
least semiclassically, i.e. there are no levels where it is violated
(cf. eq.~(\ref{eq:412a})). The second inequality marks the border in the
$(n,\ell)$-plane in Fig.~\ref{fig:2} where 
$\delta_n(\ell)$ drops to 0. For type B
motion $\ell_r+1/2\le(2E_{{n_r}\ell}/\hbar\omega_0)^{1/2}
(2\mu/\hbar\omega_0)^{1/2}$ is satisfied.
The level shifts are positive due to the repulsion provided by the
condensate. This repulsion is stronger for the lower lying levels, as one
could expect.


\subsection{Structure function:}
The dynamical structure function in the approximation we have considered is determined, up to prefactors, by just two parameters, $R_0^2$ which fixes the number of particles of a given atomic species in a given trap via $N_0=\sqrt{\hbar/m \omega_0} R_0^5/15 a$, and $k d_0$ which fixes the scattering angle via $\theta = 2 Arcsin \left [k d_0 (\lambda/4\pi) (m\omega_0/\hbar)^{1/2} \right ]$, where $\lambda$ is the wavelength of the scattered light. We shall assume that the wavelength of the light used in the scattering experiment is off-resonance but roughly given by the D-line of the alkali-metal, i.e. $\lambda \approx 590$ nm for sodium, and $\lambda \approx 800$ nm for rubidium. 

The result (\ref{eq:6.27}) for the inelastic part of the structure function
for $R_0^2/2=\mu/\hbar\omega_0=4.27$ and $kd_0=15.1$ 
is plotted in Fig.~\ref{fig:3}. For a trap with frequency $\nu_0=\omega_0/2\pi = 100$ Hz these numbers correspond to light scattering from a $^{23}Na$ condensate of $N_0 \approx 8000$ atoms with $\theta\approx 40^o$ or from a $^{87}Rb$ condensate of $N_0 \approx 4000$ atoms with $\theta \approx 126^o$. This is a typical example for light scattering with a large momentum transfer in which high-lying quasi-particle states are excited.
The plot gives as discrete points the strength of the
$\delta$-functions in eq.~(\ref{eq:6.27}), which according to our sign convention for $\omega$ appear at negative values $\omega = - E_{n_r \ell}/\hbar$ for absorption of energy. Since we consider temperature $T=0$ in the plots of Figs.~\ref{fig:3},~\ref{fig:4},~\ref{fig:6} the emission lines at positive $\omega = E_{n_r \ell}/\hbar$ are frozen out. The infinitely sharp lines described by the $\delta$-functions  are, of course, not directly
observable. Apart from the intrinsic line-widths which we did not consider in this work, they must also be
convoluted with the experimental resolution. However, in order to retain the
full information present in eq.~(\ref{eq:6.27}), we prefer not to carry out
such an arbitrary smoothing. The inelastic part of the structure function is seen to consist
of several discrete quasi-continuous rotational bands, roughly spaced by
$\hbar\omega_0$. In the region of parameter space chosen
here, these bands can be usefully labelled by the principal quantum number
$n=2n_r+\ell$ of the free trap, in terms of which the energy levels are expressed as
\begin{equation}
 E_{{n_r}\ell}=\hbar\omega_0(n+3/2)-\mu+\hbar\omega_0\delta_n(\ell)\,.
\label{eq:7.6}
\end{equation}
If $\delta_n(\ell)=0$ were satisfied each band would collapse to a single
sharp line. The semiclassical selection rule (\ref{eq:6.15}) limits the
excited bands to the domain
\begin{equation}
 \frac{1}{2}\sqrt{(kd_0)^4+2R_0^2(kd_0)^2}\ge
  \frac{E_{{n_r}\ell}}{\hbar\omega_0}\ge\frac{(kd_0)^2}{2}\,.
\label{eq:7.7}
\end{equation}
In the present case $(kd_0)^2>>2R_0^2$ we may simplify this relation somewhat
by expanding the square root and using $R_0^2=2\mu/\hbar\omega_0$ to obtain an energy interval of size $\mu$ for the excited quasi-particle states which extends above the recoil energy $\hbar^2 k^2/2 m$
\begin{equation}
 \frac{(kd_0)^2}{2}+\frac{\mu}{\hbar\omega_0}\ge
  \frac{E_{{n_r}\ell}}{\hbar\omega_0}\ge\frac{(kd_0)^2}{2}\,.
\label{eq:7.8}
\end{equation}
In the local density approximation \cite{Tom} the structure function is just a single
continuous band in the same energy interval  (and a corresponding band at negative energies related by detailed balance). It is not surprising that our treatment, being based on the discrete quasi-particle states, reveals considerably more structure, namely the bands and their discrete substructure. As can be seen from (\ref{eq:7.8}) the number of bands in Fig.~\ref{fig:3} just gives the integer part of
$\mu/\hbar\omega_0$. This rather direct physical measure of $\mu$ in units
of $\hbar\omega_0$ could be of practical experimental interest. The fine structure
of the rotational bands is also determined by the semiclassical selection
rules (\ref{eq:6.14})--(\ref{eq:6.16}). The rotational quantum number
$\ell$ increases within each band of Fig.~\ref{fig:3} 
from left to right. The strongest
line within each band corresponds to the maximum value $\ell_{max}(n)$ of $\ell$ permitted
by eq.~(\ref{eq:6.16}) for type B motion. 

We can compare these results with similar ones one may
infer from Fig.~2 of ref.~\cite{Csordas1}, where the quasi-continuous rotational bands due to high-lying quasi-particle states were first found.
The results in  \cite{Csordas1} were obtained by a much simpler 
approximation, not aiming at achieving full self-consistency, using the excited state wave functions of the
empty trap, rather than the WKB wave functions whose use underlies the
present Fig.~\ref{fig:3}.  
This implies that $\pi_r(\rho)$ in eq.~(\ref{eq:6.10}) is replaced
by the free oscillator expression
\begin{equation}
 \pi_r^0=\sqrt{\epsilon-j^2/\rho^2-\rho^2}\,.
\label{eq:7.9}
\end{equation}
Of course this leads to
much simpler closed formulas for $|M_{n,\ell}(k)|^2$ (eqs.~(5), (6) of
ref.~\cite{Csordas1}). However, with our
present self-consistent treatment we are now 
able to assess the limitations of
the simpler approximation. In fact, the result 
presented in Fig.~\ref{fig:3} differs
in two important details from that of \cite{Csordas1}:\\
(i) There is an overall shift of size $\mu$ of the energy window excited
by the transferred wave number $k$, and (ii) there is a qualitative change
in the dependence of the width of the rotational bands on the principal
quantum number which, according to Fig.~\ref{fig:3}, 
become more narrow as the
principal quantum number increases, while the opposite behavior was found
in \cite{Csordas1}. Both differences 
have their origin in a change in the semiclassical
selection rules (\ref{eq:6.14})--(\ref{eq:6.16}), which depend rather
sensitively on approximations in the phases of the wave functions inside the
condensate. 
The first
shortcoming of the approximation in 
\cite{Csordas1} can be easily eliminated
by using simply $E_n^{{\it osc}}=\hbar\omega_0(n+3/2)-\mu$ rather than the free trap
energies  in the expressions for the free trap
eigenfunctions. The second is not as
easily avoided, however. It simply shows
that the influence of the interaction with the condensate on the spatial
dependence of the phase of the high lying states cannot be neglected as long
as the widths $\Delta_n$ of the rotational bands given by $\Delta_n=\delta_{n} (0)-\delta_{n} (\ell_{max}(n))$ 
with
\begin{equation}
 \ell_{\max}(n)+1/2 =
  \sqrt{2}\left((kd_0)^4/4+(kd_0)^2 \mu/\hbar\omega_0-
   (E_n^{{\it osc}}/\hbar\omega_0)^2\right)^{1/2}
\label{eq:7.10}
\end{equation}
are not themselves negligible. We conclude that for a proper treatment of the rotational bands
the present treatment is
indispensable. 

The spectroscopic resolution of the quasi-continuous bands requires a resolution better than the widths $\Delta_n$, which may be difficult to attain in the frequency domain, but could be feasible in real-time experiments extended over a time-scale of $\approx 10$ periods $2\pi/\omega_0$.  
In real time the width $\Delta_n$ would turn up as collapse rate $\gamma_n=\Delta_n$ determining the dephasing of the band with principal quantum number $n$. Due to the discrete substructure of the bands this dephasing should in principle be reversible. However it appears unlikely that the strict phase coherence over a time-interval of the order of $\ell_{max}(n)/\Delta_n$, required to see such a rephasing, could be achieved experimentally. For a discussion of collapses and revivals of low-lying collective modes see \cite{p} \cite{GCW}.

Let us now look at the dynamical structure function also for smaller
momentum transfer and for somewhat larger condensates. In fig.~\ref{fig:4}
we have chosen the parameters $R^2_0=11.24$, $kd_0=3$ corresponding to,
$N_0=6000$, $\theta=21^\circ$ for $^{87}Rb$ and to $N_0=12000$,
$\theta=7.7^\circ$ for $^{23}Na$,
if the same choices are made for $\nu_0$ and $\lambda$ as before.
The changes in the scale in fig.~\ref{fig:4} in comparison with
fig.~\ref{fig:3}
should be noted. In fig.~\ref{fig:4} we have joined spectral lines with
equal principal quantum number $n=2n_r+\ell$ by straight lines. This
is usefull because the quantum number $n$ organizes the spectral lines according
to multiplets of different strengths. However, differently to the case
of high lying states, where $n$ labelled well-separated quasi-continuous
bands, the different multiplets now overlap at least partially
in their energy range. The strongest multiplet in fig.~\ref{fig:4}
corresponds to $n=9$, the next to $n=8$, and the third and weakest one
shown to $n=7$. There are actually still weaker lines with $n\le 6$ which we have
suppressed. All the lines
satisfy the semiclassical selection rule (\ref{eq:6.20a}). It is plotted
for this case in fig.~\ref{fig:5} in the (angular momentum, energy)-plane,
together with the semiclassical energy levels. All levels within the
curved triangular section of the plane satisfy the selection rule.
In this plot multiplets of fixed $n=2n_r+\ell$ lie on lines roughly
parallel to the upper border of the triangular shape, the higher lying
lines corresponding to higher values of $n$. On this upper border the
square-root in the denominator of the expression (\ref{eq:6.27}) for
$S^{(1)}(\bbox{k},\omega)$ vanishes, leading to a scattering singularity,
while on the lower border the nominator of this expression vanishes,
leading to a suppression of scattering. This observation explains
qualitatively the dependence of the scattering cross-section on $n$.
The multiplet with $n=9$ consists only of the two lines for $\ell=1$,
$€\ell=3$. The remaining $\ell=5$, $\ell=7$, $\ell=9$ members of this multiplet
have smaller energy but lie above the triangular region in fig.~\ref{fig:5}.
This highest lying multiplet of fig.~\ref{fig:5} is therefore rather narrow
and well separated in energy from the other multiplets. It lies entirely above the chemical
potential $\mu$, i.e. its lines in fig.~\ref{fig:4} lie at
$\omega/\omega_0<\mu/\hbar \omega_0 =-5.62$. With these features this multiplet bears some resemblance to the well-separated bands of fig.~\ref{fig:3}.
In contrast to this the two lower lying and weaker multiplets in fig.~\ref{fig:4}
overlap completely with each other and extend over energy levels
ranging from values larger to values smaller than $\mu$ and already resemble
the low lying multiplets extending down to the hydrodynamic
regime. This can be seen from fig.~\ref{fig:6} where we have chosen
$R_0^2=26.2$, $kd_0=2$ corresponding to $N_0=50000$, $\theta=14^\circ$
for $^{87}Rb$ and $N_0=10^5$, $\theta=5.2^\circ$ for $^{23}Na$, again
assuming a trap of $\nu_0=100$~Hz and off-resonance scattering at
wavelengths roughly equal to the respective $D$-lines. The multiplets
are still organized according to their strengths by the principal
quantum number $n$. However, now these multiplets are completely
overlapping, much as the two lower lying ones in fig.~\ref{fig:4}.
For the multiplets shown in fig.~\ref{fig:6} $n$ ranges from $n=8$
for the strongest multiplet to $n=6$ for the weakest one shown in the
figure, but again we have suppressed in the plot still weaker lines
corresponding to multiplets with smaller values of $n$. The dependence
of the scattering cross-section on $n$, and within a given multiplet
on $\ell$, can be understood similarly to our discussion of
fig.~\ref{fig:4} by using the asymptotic expression (\ref{eq:6.34}) for
$S^{(1)}(\bbox{k},\omega)$. Taken together the plots
fig.~\ref{fig:3}--fig.~\ref{fig:6} show that a measurement of
$S(\bbox{k},\omega)$ over a range of scattering angles $\theta$ in principle
can give a very detailed information on the quasiparticle levels in
any energy interval selected by the choice of $k$ according to
eq.~(\ref{eq:6.20a}). 

\section{Conclusions}
\label{sec:conclusion}
In the present paper we have solved the Bogoliubov equations for the  
elementary excitations of a Bose-condensed gas of atoms trapped in an  
isotropic harmonic potential in WKB-approximation. The underlying classical  
dynamics is integrable in this case due to the rotational symmetry. The  
quasi-particle energies in the trap have been obtained from a  
Bohr-Sommerfeld quantization rule. Corrections to that rule due to the  
non-differentiabilty of the condensate wave function in Thomas-Fermi  
approximation have been discussed in \cite{Csordas2}. The mode-functions,  
including the Bogoliubov amplitudes $v_j$, have been obtained in the  
classically allowed regime in the WKB-form. Our results interpolate between  
the low-energy regime $E/\mu \to 0$, where asymptotically exact results within the Bogoliubov approximation are  
available \cite{6b} and the high-energy regime $\mu/E \to 0$, where  
perturbation theory can be used. As an application we have given a self-consistent calculation of the fully resolved 
dynamical structure function $S(k,\omega)$,
which could be  measured in off-resonant light  
scattering, in a 
large energy domain extending to all but the  
lowest lying levels.

\section*{Acknowledgements}

One of us (R.G.) wishes to acknowledge support by the Deutsche
Forschungsgemeinschaft through the Sonderforschungsbereich 237
``Unordnung und gro{\ss}e Fluktuationen''.
This work has been supported by the project of the Hungarian Academy
of Sciences and the Deutsche Forschungsgemeinschaft under grant No. 95 and
has also been supported within the framework of the
German-Hungarian Scientific and Technological Cooperation under
Project No. 62.
Two of us (A. Cs., P. Sz.) would like to acknowledge support
by the Hungarian Academy of Sciences under grant No. AKP 96-12/14.
The work has been partially supported by the Hungarian National
Scientific Research Foundation under grant Nos. OTKA T017493,
F020094 and by the Ministry of Culture and Education of Hungary
under grant No. FKFP0159/1997.

\appendix
\section{Evaluation of the action integral}
\label{sec:appa}

Here we summarize the essential steps for the evaluation of the integral in
eq.~(\ref{eq:4.29}). The integral is splitted into two parts, the first part
$I_<$
taken over $\rho$ from $\rho_<$ to 1, the second part $I_>$ taken over
$\rho$ from 1 to $\rho_>$. Let us consider the evaluation of $I_<$ first.
The substitution
\begin{eqnarray}
 \rho &=& \sqrt{1-\frac{2\epsilon t}{1-t^2}}\nonumber\\
 d\rho &=& -\epsilon\frac{(1+t^2)}{(1-t^2)^{3/2}(1-t^2-2\epsilon t)^{1/2}}dt
\label{eq:a1}
\end{eqnarray}
transforms $I_<$ into an integral over $t$ from
\begin{equation}
 t=0\quad \mbox{\rm to}\quad t=t_{\max}=
  \frac{-(\epsilon^2+j^2)+\epsilon\sqrt{1+\epsilon^2+2j^2}}{\epsilon+j^2}\,.
\label{eq:a2}
\end{equation}
A partial fraction expansion of the integrand results in the explicit form
\begin{eqnarray}
 I_<=&& \frac{\hbar R_0^2}{\pi}\epsilon\sqrt{\epsilon^2+j^2}\int_0^{t_{\max}}
  dt\sqrt{\frac{\epsilon^2(1+\epsilon^2+2j^2)}{(\epsilon+j^2)^2}-
  \left(t+\frac{\epsilon^2+j^2}{\epsilon+j^2}\right)}\nonumber\\
  &&\quad\Bigg\{\frac{\sqrt{\epsilon^2+1}+\epsilon-1}{4\epsilon^2}\,
     \frac{1}{\sqrt{\epsilon^2+1}-\epsilon-t}+
      \frac{\sqrt{\epsilon^2+1}-\epsilon+1}{4\epsilon^2}\,
       \frac{1}{\sqrt{\epsilon^2+1}+\epsilon+t}\nonumber\\
  &&\qquad -\frac{1}{4\epsilon}\,\frac{1}{1-t}+\frac{\epsilon-2}{4\epsilon^2}
      \,\frac{1}{1+t}+\frac{1}{2\epsilon(1+t)^2}\Bigg\}
\label{eq:a3}
\end{eqnarray}

A further substitution
\begin{equation}
 t=\frac{\epsilon\sqrt{1+\epsilon^2+2j^2}}{\epsilon+j^2}\sin\varphi-
  \frac{\epsilon^2+j^2}{\epsilon+j^2}
\label{eq:a4}
\end{equation}
and treating the curly bracket in eq.~(\ref{eq:a3}) term by term reduces the
integral to the form
\begin{eqnarray}
 I_<=\frac{\hbar R_0^2}{\pi}\Bigg\{&&\sum_{i=1}^3C_iJ_>(A_1,a_i,b_1)+
    C_4J_<(A_1,a_4,b_1)\nonumber\\
    && + C_5K_<(A_1,a_4,b_1)\Bigg\}
\label{eq:a5}
\end{eqnarray}
with the functions $J_<$, $J_>$, $K_<$ defined in
eq.~(\ref{eq:4.30})--(\ref{eq:4.31}) and the parameters and coefficients given
in eqs.~(\ref{eq:4.33})--(\ref{eq:4.45}).

Turning to the second part $I_>$ of $I_r$ we may apply a similar sequence of
substitutions resulting in
\begin{equation}
 \rho=\sqrt{\frac{\epsilon+1}{2}+\sqrt{\left(\frac{\epsilon+1}{2}\right)^2
   -j^2}\sin\varphi}
\label{eq:a6}
\end{equation}
which brings $\mbox{\rm I}_>$ to the form
\[
 I_>=\frac{\hbar R_0^2}{\pi}C_6J_>(A_2,a_5,b_2)
\]
with the parameters and coefficients given in
eqs.~(\ref{eq:4.33})--(\ref{eq:4.45}). From $I_r=I_<+I_>$ we obtain the result
given in eq.~(\ref{eq:4.32}).

\section{Normalization of the WKB wave function}
\label{sec:appb}

Here we summarize the necessary steps in the evaluation of the integral in
eq.~(\ref{eq:5.3}). For type A motion the radial period is simply
$\pi/\omega_0$. To evaluate eq.~(\ref{eq:5.3}) also for type B motion we
introduce scaled variables like in eq.~(\ref{eq:4.7}), and the integral
is splitted in a part $I_1$ from $\rho_<$ to 1 and a part $I_2$ from 1 to
$\rho_>$.
\begin{eqnarray}
\label{eq:b1}
I_1 &=& \frac{1}{m\omega_0}\int_{\rho_<}^1d\rho\,
 \frac{\epsilon}{\sqrt{\epsilon^2+(1-\rho^2)^2}}\,
 \frac{1}{\sqrt{-\left(\frac{j^2}{\rho^2}+1-\rho^2\right)+
  \sqrt{\epsilon^2+(1-\rho^2)^2}}}\\
I_2 &=& \frac{1}{m\omega_0}\int_1^{\rho_>}d\rho\,
  \frac{1}{\sqrt{\epsilon+1-\frac{j^2}{\rho^2}-\rho^2}}\,.
\label{eq:b2}
\end{eqnarray}
A series of substitutions amounting to
\begin{eqnarray}
\rho^2 &=& \frac{1-t^2-2\epsilon t}{1-t^2}\nonumber\\
 t &=& \left[\epsilon\sqrt{1+\epsilon^2+2j^2}
  \sin\varphi-\epsilon^2-j^2\right]/(\epsilon+j^2)
\label{eq:b3}
\end{eqnarray}
puts $I_1$ into the form
\begin{equation}
 I_1 = \frac{\epsilon\sqrt{\epsilon+j^2}}{m\omega_0}\int_{A_1}^{\pi/2}
  \frac{d\varphi}
   {\epsilon+\epsilon^2+2j^2-\epsilon\sqrt{1+\epsilon^2+2j^2}\sin\varphi}
\label{eq:b4}
\end{equation}
where $A_1$ is defined in (\ref{eq:4.33}). The integral is now elementary
and given by
\begin{equation}
 I_1=\frac{2\epsilon}{\sqrt{2(\epsilon^2+2j^2)}}\,\frac{\beta}{m\omega_0}
\label{eq:b5}
\end{equation}
with $\beta$ given by eq.~(\ref{eq:5.5}). Here we used the addition theorem
for arctan to simplify the final expression.

$I_2$ is much simpler to evaluate. The substitution
\begin{equation}
 \rho^2=\frac{\epsilon+1}{2}+\sqrt{\left(\frac{\epsilon+1}{2}\right)^2-j^2}
  \sin\varphi
\label{eq:b6}
\end{equation}
reduces it to the form
\begin{equation}
 I_2=\frac{1}{2m\omega_0}\int_{A_2}^{\pi/2}d\varphi
\label{eq:b7}
\end{equation}
where $A_2$ is defined in eq.~(\ref{eq:4.34}). After some simplification the
result takes the form
\begin{equation}
 I_2=\frac{1}{2m\omega_0}\left(\frac{\pi}{2}-\alpha\right)
\label{eq:b8}
\end{equation}
with $\alpha$ given by eq.~(\ref{eq:5.5}). Combining the results
(\ref{eq:b5}), (\ref{eq:b8}) with eq.~(\ref{eq:5.3}) we obtain
eq.~(\ref{eq:5.4}).


\begin{figure}
\caption{\label{fig:1}Regions of type A motion and type B motion in the
scaled angular momentum-energy $(j-\epsilon)$-plane. In region D there is no lower turning
point of the radial motion, i.e., in the semiclassical quantization region D is forbidden.}
\end{figure}

\begin{figure}
\caption{\label{fig:2}
Difference $\delta =\delta_n(\ell)$ of quasi-particle energy levels,
in units of $\hbar \omega_0$, 
to the levels
of the free trap. The chemical potential was chosen to be
$\mu=4.27 \hbar\omega_0$.}
\end{figure}

\begin{figure}
\caption{\label{fig:3}
 Dynamic structure function $S^{(1)}(k,\omega)$ in units of $d_0/a\hbar\omega_0$ for
$\mu/\hbar\omega_0= 4.27$, $kd_0=15.1$ as a function of $\omega$
in units of $\omega_0$. Plotted are the strengths of the
Dirac-delta peaks.}
\end{figure}

\begin{figure}
\caption{\label{fig:4}
 Dynamic structure function $S^{(1)}(k,\omega)$ in units of $d_0/a\hbar\omega_0$ for
$\mu/\hbar\omega_0= 5.62$, $kd_0=3$ as a function of $\omega$
in units of $\omega_0$. Plotted are the strengths of the
Dirac-delta peaks. Peaks with equal principal quantum number $n=2 n_r +\ell$ are connected by straight lines, with $n=9,8,7$ from top to bottom.}
\end{figure}

\begin{figure}
\caption{\label{fig:5}
 Semiclassical selection rule \ref{eq:6.20a} and semiclassical energy levels $E=E_{n_r \ell}$ in the ($\ell , E$)-plane for $\mu/\hbar\omega_0 = 5.62$, $kd_0=3$.}
\end{figure}

\begin{figure}
\caption{\label{fig:6}
 Dynamic structure function $S^{(1)}(k,\omega)$ in units of $d_0/a\hbar\omega_0$ for
$\mu/\hbar\omega_0= 13.1$, $kd_0=2$ as a function of $\omega$ in units of $\omega_0$. Plotted are the strengths of the
Dirac-delta peaks. Peaks with equal principal quantum number $n=2 n_r +\ell$ are connected by straight lines, with $n=8,7,6$ from top to bottom.}
\end{figure}

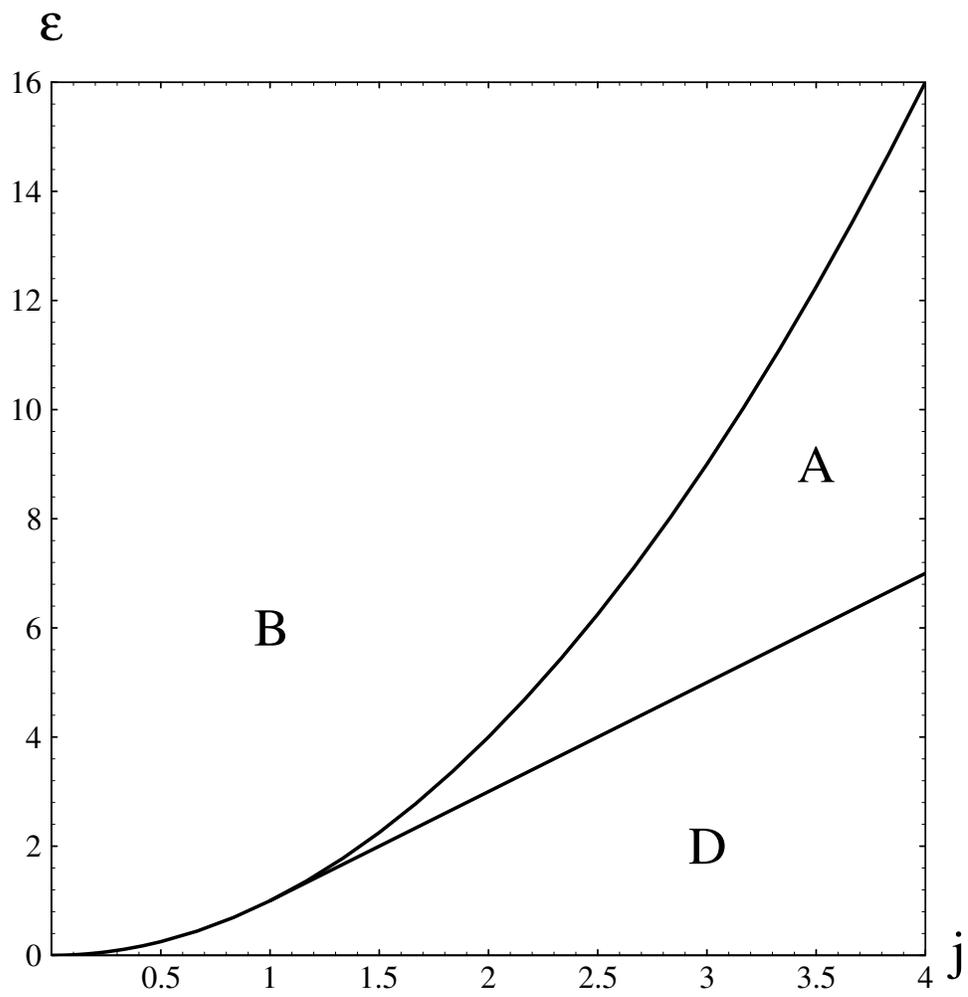
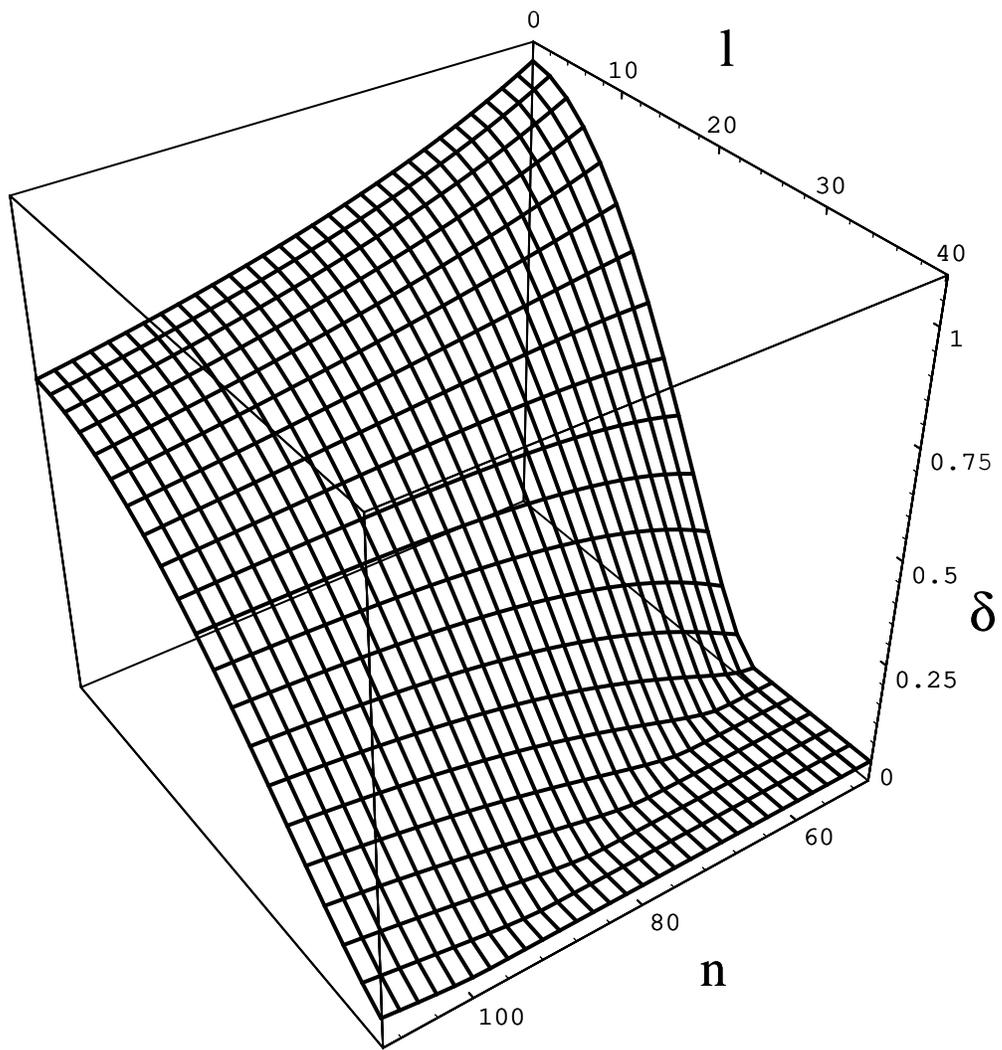
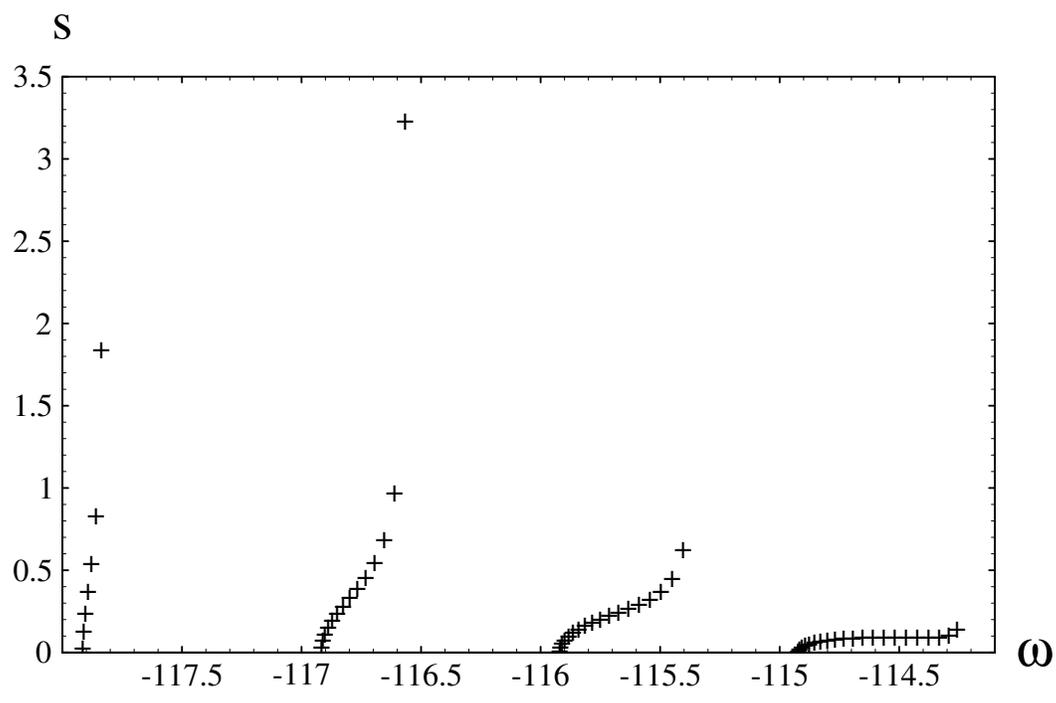
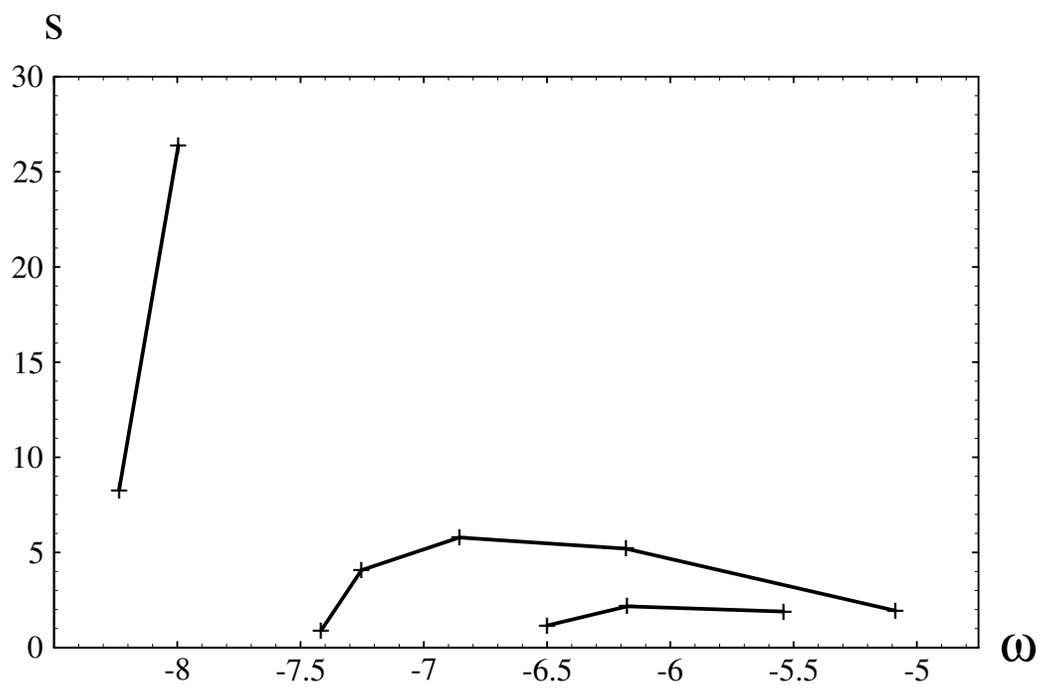
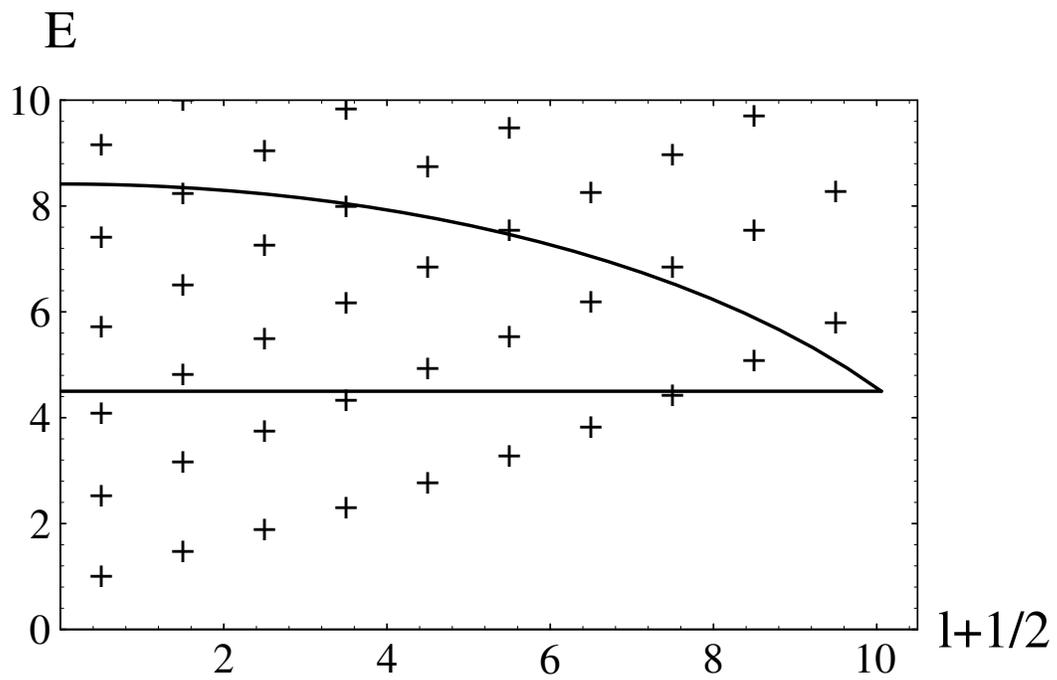
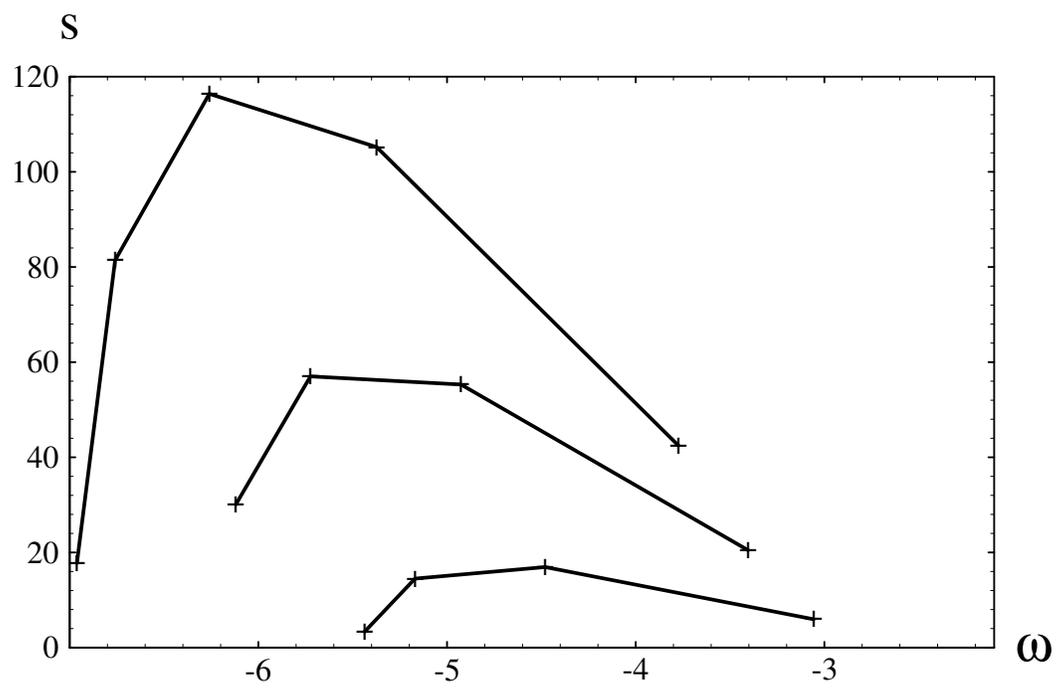
\end{document}